\documentclass[prl,
secnumarabic,twocolumn,twoside,a4paper,nofootinbib,nobibnotes,showpacs,amsmath,amssymb,floatfix%
,italian,frenchb,german,
swedish,british%
]{revtex4}

\newcommand{\usetxfonts}{\usepackage{txfonts}}

\newcommand*\savesymbol[1]{%
  \expandafter\let\csname orig#1\expandafter\endcsname\csname#1\endcsname
  \expandafter\let\csname #1\endcsname\relax
}

\newcommand*\restoresymbol[2]{%
  \expandafter\global\expandafter\let\csname#1#2\expandafter\endcsname%
    \csname#2\endcsname
  \expandafter\global\expandafter\let\csname#2\expandafter\endcsname%
    \csname orig#2\endcsname
}

\usepackage[T1]{fontenc} 
\usepackage{textcomp}
\usepackage{amsxtra}

\usepackage[
]{babel} 
\newcommand{\langfrench}{\foreignlanguage{french}}

\newcommand{\langlatin}{\foreignlanguage{latin}}

\newcommand{\latin}[1]{\emph{\langlatin{#1}}}
\newcommand{\french}[1]{\emph{\langfrench{#1}}}

\savesymbol{qedsymbol}

\usepackage{amsthm}

\restoresymbol{ams}{qedsymbol}
\newcommand{\QEE}
{\amsqedsymbol}

\usepackage[dvipdf]{graphicx}
\usepackage{url}

\providecommand{\usetxfonts}{}
\usetxfonts

\DeclareFontFamily{U}{egreek}{\skewchar\font'177}%
\DeclareFontShape{U}{egreek}{m}{n}{<-6>s*[1]eurm5<6-8>s*[1]eurm7 <8->s*[1]eurm10}{}%
\DeclareFontShape{U}{egreek}{m}{it}{<->s*[1]eurmo10}{}%
\DeclareFontShape{U}{egreek}{b}{n}{<-6>s*[1]eurb5 <6-8>s*[1]eurb7 <8->s*[1]eurb10}{}%
\DeclareFontShape{U}{egreek}{b}{it}{<->s*[1]eurbo10}{}

\DeclareFontFamily{U}{egreek}{\skewchar\font'177}%
\DeclareFontShape{U}{egreek}{m}{n}{<-6>s*[1]eurm5<6-8>s*[1]eurm7 <8->s*[1]eurm10}{}%
\DeclareFontShape{U}{egreek}{m}{it}{<->s*[1]eurmo10}{}%
\DeclareFontShape{U}{egreek}{b}{n}{<-6>s*[1]eurb5 <6-8>s*[1]eurb7 <8->s*[1]eurb10}{}%
\DeclareFontShape{U}{egreek}{b}{it}{<->s*[1]eurbo10}{}%
\DeclareSymbolFont{egreeki}{U}{egreek}{m}{it}%
\SetSymbolFont{egreeki}{bold}{U}{egreek}{b}{it}
\DeclareSymbolFontAlphabet{\mathegri}{egreeki}%
\DeclareMathSymbol{\epartial}{\mathalpha}{egreeki}{"40}

\DeclareMathSymbol{\ealpha}{\mathalpha}{egreeki}{"0B}
\DeclareMathSymbol{\ebeta}{\mathalpha}{egreeki}{"0C}
\DeclareMathSymbol{\egamma}{\mathalpha}{egreeki}{"0D}
\DeclareMathSymbol{\edelta}{\mathalpha}{egreeki}{"0E}
\DeclareMathSymbol{\eepsilon}{\mathalpha}{egreeki}{"0F}
\DeclareMathSymbol{\ezeta}{\mathalpha}{egreeki}{"10}
\DeclareMathSymbol{\eeta}{\mathalpha}{egreeki}{"11}
\DeclareMathSymbol{\etheta}{\mathalpha}{egreeki}{"12}
\DeclareMathSymbol{\eiota}{\mathalpha}{egreeki}{"13}
\DeclareMathSymbol{\ekappa}{\mathalpha}{egreeki}{"14}
\DeclareMathSymbol{\elambda}{\mathalpha}{egreeki}{"15}
\DeclareMathSymbol{\emu}{\mathalpha}{egreeki}{"16}
\DeclareMathSymbol{\enu}{\mathalpha}{egreeki}{"17}
\DeclareMathSymbol{\exi}{\mathalpha}{egreeki}{"18}
\DeclareMathSymbol{\eomicron}{\mathalpha}{egreeki}{"6F}
\DeclareMathSymbol{\epi}{\mathalpha}{egreeki}{"19}
\DeclareMathSymbol{\erho}{\mathalpha}{egreeki}{"1A}
\DeclareMathSymbol{\esigma}{\mathalpha}{egreeki}{"1B}
\DeclareMathSymbol{\etau}{\mathalpha}{egreeki}{"1C}
\DeclareMathSymbol{\eupsilon}{\mathalpha}{egreeki}{"1D}
\DeclareMathSymbol{\ephi}{\mathalpha}{egreeki}{"1E}
\DeclareMathSymbol{\echi}{\mathalpha}{egreeki}{"1F}
\DeclareMathSymbol{\epsi}{\mathalpha}{egreeki}{"20}
\DeclareMathSymbol{\eomega}{\mathalpha}{egreeki}{"21}
\DeclareMathSymbol{\evarepsilon}{\mathalpha}{egreeki}{"22}
\DeclareMathSymbol{\evartheta}{\mathalpha}{egreeki}{"23}
\DeclareMathSymbol{\evarpi}{\mathalpha}{egreeki}{"24}
\let\evarrho\erho 
\let\evarsigma\esigma
\let\evarkappa\ekappa
\DeclareMathSymbol{\evarphi}{\mathalpha}{egreeki}{"27}
\DeclareMathSymbol{\evarAlpha}{\mathalpha}{egreeki}{"41}
\DeclareMathSymbol{\evarBeta}{\mathalpha}{egreeki}{"42}
\DeclareMathSymbol{\evarGamma}{\mathalpha}{egreeki}{"00}
\DeclareMathSymbol{\evarDelta}{\mathalpha}{egreeki}{"01}
\DeclareMathSymbol{\evarEpsilon}{\mathalpha}{egreeki}{"45}
\DeclareMathSymbol{\evarZeta}{\mathalpha}{egreeki}{"5A}
\DeclareMathSymbol{\evarEta}{\mathalpha}{egreeki}{"48}
\DeclareMathSymbol{\evarTheta}{\mathalpha}{egreeki}{"02}
\DeclareMathSymbol{\evarIota}{\mathalpha}{egreeki}{"49}
\DeclareMathSymbol{\evarKappa}{\mathalpha}{egreeki}{"4B}
\DeclareMathSymbol{\evarLambda}{\mathalpha}{egreeki}{"03}
\DeclareMathSymbol{\evarMu}{\mathalpha}{egreeki}{"4D}
\DeclareMathSymbol{\evarNu}{\mathalpha}{egreeki}{"4E}
\DeclareMathSymbol{\evarXi}{\mathalpha}{egreeki}{"04}
\DeclareMathSymbol{\evarOmicron}{\mathalpha}{egreeki}{"4F}
\DeclareMathSymbol{\evarPi}{\mathalpha}{egreeki}{"05}
\DeclareMathSymbol{\evarRho}{\mathalpha}{egreeki}{"50}
\DeclareMathSymbol{\evarSigma}{\mathalpha}{egreeki}{"06}
\DeclareMathSymbol{\evarTau}{\mathalpha}{egreeki}{"54}
\DeclareMathSymbol{\evarUpsilon}{\mathalpha}{egreeki}{"07}
\DeclareMathSymbol{\evarPhi}{\mathalpha}{egreeki}{"08}
\DeclareMathSymbol{\evarChi}{\mathalpha}{egreeki}{"58}
\DeclareMathSymbol{\evarPsi}{\mathalpha}{egreeki}{"09}
\DeclareMathSymbol{\evarOmega}{\mathalpha}{egreeki}{"0A} 
\DeclareFontFamily{U}{egreek}{\skewchar\font'177}%
\DeclareFontShape{U}{egreek}{m}{n}{<-6>s*[1]eurm5<6-8>s*[1]eurm7 <8->s*[1]eurm10}{}%
\DeclareFontShape{U}{egreek}{m}{it}{<->s*[1]eurmo10}{}%
\DeclareFontShape{U}{egreek}{b}{n}{<-6>s*[1]eurb5 <6-8>s*[1]eurb7 <8->s*[1]eurb10}{}%
\DeclareFontShape{U}{egreek}{b}{it}{<->s*[1]eurbo10}{}%
\DeclareSymbolFont{egreekr}{U}{egreek}{m}{n}%
\SetSymbolFont{egreekr}{bold}{U}{egreek}{b}{n}
\DeclareSymbolFontAlphabet{\mathegrr}{egreekr}%
\DeclareMathSymbol{\epartialup}{\mathalpha}{egreekr}{"40}

\DeclareMathSymbol{\ealphaup}{\mathalpha}{egreekr}{"0B}
\DeclareMathSymbol{\ebetaup}{\mathalpha}{egreekr}{"0C}
\DeclareMathSymbol{\egammaup}{\mathalpha}{egreekr}{"0D}
\DeclareMathSymbol{\edeltaup}{\mathalpha}{egreekr}{"0E}
\DeclareMathSymbol{\eepsilonup}{\mathalpha}{egreekr}{"0F}
\DeclareMathSymbol{\ezetaup}{\mathalpha}{egreekr}{"10}
\DeclareMathSymbol{\eetaup}{\mathalpha}{egreekr}{"11}
\DeclareMathSymbol{\ethetaup}{\mathalpha}{egreekr}{"12}
\DeclareMathSymbol{\eiotaup}{\mathalpha}{egreekr}{"13}
\DeclareMathSymbol{\ekappaup}{\mathalpha}{egreekr}{"14}
\DeclareMathSymbol{\elambdaup}{\mathalpha}{egreekr}{"15}
\DeclareMathSymbol{\emuup}{\mathalpha}{egreekr}{"16}
\DeclareMathSymbol{\enuup}{\mathalpha}{egreekr}{"17}
\DeclareMathSymbol{\exiup}{\mathalpha}{egreekr}{"18}
\DeclareMathSymbol{\eomicronup}{\mathalpha}{egreekr}{"6F}
\DeclareMathSymbol{\epiup}{\mathalpha}{egreekr}{"19}
\DeclareMathSymbol{\erhoup}{\mathalpha}{egreekr}{"1A}
\DeclareMathSymbol{\esigmaup}{\mathalpha}{egreekr}{"1B}
\DeclareMathSymbol{\etauup}{\mathalpha}{egreekr}{"1C}
\DeclareMathSymbol{\eupsilonup}{\mathalpha}{egreekr}{"1D}
\DeclareMathSymbol{\ephiup}{\mathalpha}{egreekr}{"1E}
\DeclareMathSymbol{\echiup}{\mathalpha}{egreekr}{"1F}
\DeclareMathSymbol{\epsiup}{\mathalpha}{egreekr}{"20}
\DeclareMathSymbol{\eomegaup}{\mathalpha}{egreekr}{"21}
\DeclareMathSymbol{\evarepsilonup}{\mathalpha}{egreekr}{"22}
\DeclareMathSymbol{\evarthetaup}{\mathalpha}{egreekr}{"23}
\DeclareMathSymbol{\evarpiup}{\mathalpha}{egreekr}{"24}
\let\evarrhoup\erhoup 
\let\evarsigmaup\esigmaup
\let\evarkappaup\ekappaup
\DeclareMathSymbol{\evarphiup}{\mathalpha}{egreekr}{"27}
\DeclareMathSymbol{\eAlpha}{\mathalpha}{egreekr}{"41}
\DeclareMathSymbol{\eBeta}{\mathalpha}{egreekr}{"42}
\DeclareMathSymbol{\eGamma}{\mathalpha}{egreekr}{"00}
\DeclareMathSymbol{\eDelta}{\mathalpha}{egreekr}{"01}
\DeclareMathSymbol{\eEpsilon}{\mathalpha}{egreekr}{"45}
\DeclareMathSymbol{\eZeta}{\mathalpha}{egreekr}{"5A}
\DeclareMathSymbol{\eEta}{\mathalpha}{egreekr}{"48}
\DeclareMathSymbol{\eTheta}{\mathalpha}{egreekr}{"02}
\DeclareMathSymbol{\eIota}{\mathalpha}{egreekr}{"49}
\DeclareMathSymbol{\eKappa}{\mathalpha}{egreekr}{"4B}
\DeclareMathSymbol{\eLambda}{\mathalpha}{egreekr}{"03}
\DeclareMathSymbol{\eMu}{\mathalpha}{egreekr}{"4D}
\DeclareMathSymbol{\eNu}{\mathalpha}{egreekr}{"4E}
\DeclareMathSymbol{\eXi}{\mathalpha}{egreekr}{"04}
\DeclareMathSymbol{\eOmicron}{\mathalpha}{egreekr}{"4F}
\DeclareMathSymbol{\ePi}{\mathalpha}{egreekr}{"05}
\DeclareMathSymbol{\eRho}{\mathalpha}{egreekr}{"50}
\DeclareMathSymbol{\eSigma}{\mathalpha}{egreekr}{"06}
\DeclareMathSymbol{\eTau}{\mathalpha}{egreekr}{"54}
\DeclareMathSymbol{\eUpsilon}{\mathalpha}{egreekr}{"07}
\DeclareMathSymbol{\ePhi}{\mathalpha}{egreekr}{"08}
\DeclareMathSymbol{\eChi}{\mathalpha}{egreekr}{"58}
\DeclareMathSymbol{\ePsi}{\mathalpha}{egreekr}{"09}
\DeclareMathSymbol{\eOmega}{\mathalpha}{egreekr}{"0A}

\newcommand{\alleugreek}{%
\let\alpha\ealpha
\let\beta\ebeta
\let\gamma\egamma
\let\delta\edelta
\let\epsilon\eepsilon
\let\zeta\ezeta
\let\eta\eeta
\let\theta\etheta
\let\iota\eiota
\let\kappa\ekappa
\let\lambda\elambda
\let\mu\emu
\let\nu\enu
\let\xi\exi
\let\omicron\eomicron
\let\pi\epi
\let\rho\erho
\let\sigma\esigma
\let\tau\etau
\let\upsilon\eupsilon
\let\phi\ephi
\let\chi\echi
\let\psi\epsi
\let\omega\eomega
\let\varepsilon\evarepsilon
\let\vartheta\evartheta
\let\varpi\evarpi
\let\varrho\evarrho 
\let\varsigma\evarsigma
\let\varkappa\evarkappa
\let\varphi\evarphi
\let\varAlpha\evarAlpha
\let\varBeta\evarBeta
\let\varGamma\evarGamma
\let\varDelta\evarDelta
\let\varEpsilon\evarEpsilon
\let\varZeta\evarZeta
\let\varEta\evarEta
\let\varTheta\evarTheta
\let\varIota\evarIota
\let\varKappa\evarKappa
\let\varLambda\evarLambda
\let\varMu\evarMu
\let\varNu\evarNu
\let\varXi\evarXi
\let\varOmicron\evarOmicron
\let\varPi\evarPi
\let\varRho\evarRho
\let\varSigma\evarSigma
\let\varTau\evarTau
\let\varUpsilon\evarUpsilon
\let\varPhi\evarPhi
\let\varChi\evarChi
\let\varPsi\evarPsi
\let\varOmega\evarOmega 
\let\alphaup\ealphaup
\let\betaup\ebetaup
\let\gammaup\egammaup
\let\deltaup\edeltaup
\let\epsilonup\eepsilonup
\let\zetaup\ezetaup
\let\etaup\eetaup
\let\thetaup\ethetaup
\let\iotaup\eiotaup
\let\kappaup\ekappaup
\let\lambdaup\elambdaup
\let\muup\emuup
\let\nuup\enuup
\let\xiup\exiup
\let\omicronup\eomicronup
\let\piup\epiup
\let\rhoup\erhoup
\let\sigmaup\esigmaup
\let\tauup\etauup
\let\upsilonup\eupsilonup
\let\phiup\ephiup
\let\chiup\echiup
\let\psiup\epsiup
\let\omegaup\eomegaup
\let\varepsilonup\evarepsilonup
\let\varthetaup\evarthetaup
\let\varpiup\evarpiup
\let\varrhoup\evarrhoup 
\let\varsigmaup\evarsigmaup
\let\varkappaup\evarkappaup
\let\varphiup\evarphiup
\let\Alpha\eAlpha
\let\Beta\eBeta
\let\Gamma\eGamma
\let\Delta\eDelta
\let\Epsilon\eEpsilon
\let\Zeta\eZeta
\let\Eta\eEta
\let\Theta\eTheta
\let\Iota\eIota
\let\Kappa\eKappa
\let\Lambda\eLambda
\let\Mu\eMu
\let\Nu\eNu
\let\Xi\eXi
\let\Omicron\eOmicron
\let\Pi\ePi
\let\Rho\eRho
\let\Sigma\eSigma
\let\Tau\eTau
\let\Upsilon\eUpsilon
\let\Phi\ePhi
\let\Chi\eChi
\let\Psi\ePsi
\let\Omega\eOmega
}

\usepackage{bm}
\providecommand{\piup}{\epiup}
\providecommand{\deltaup}{\edeltaup}
\providecommand{\zetaup}{\ezetaup}
\providecommand{\varepsilonup}{\evarepsilonup}

\providecommand{\coloneqq}{\mathrel{\mathop:}=}
\providecommand{\eqqcolon}{=\mathrel{\mathop:}}

\newcommand{\pu}{\piup}
\newcommand{\delt}{\deltaup}
\newcommand{\I}{\mathrm{i}}
\newcommand{\e}{\mathrm{e}}
\newcommand{\di}{\mathrm{d}}

\newcommand{\ZZ}{\mathbb{Z}}

\newcommand{\RR}{\mathbb{R}}

\DeclareMathOperator{\tr}{tr}
\newcommand{\herm}{{}^{\mathord{\dagger}}}

\newcommand{\defd}{\coloneqq}
\newcommand{\defs}{\eqqcolon}

\newcommand{\Lor}{\bigvee}

\newcommand{\cond}{\mathpunct{|}}
\newcommand{\st}{\mathpunct{|}}
\newcommand{\with}{\colon}

\newcommand{\corr}{\mathrel{\hat{=}}}
\renewcommand{\le}{\leqslant}
\renewcommand{\ge}{\geqslant}
\DeclareMathDelimiter{\lclose}{\mathopen}{operators}{"5B}{largesymbols}{"02}
\DeclareMathDelimiter{\rclose}{\mathclose}{operators}{"5D}{largesymbols}{"03}
\DeclareMathDelimiter{\lopen}{\mathopen}{operators}{"5D}{largesymbols}{"03}
\DeclareMathDelimiter{\ropen}{\mathclose}{operators}{"5B}{largesymbols}{"02}
\newcommand{\clcl}[1]{\lclose#1\rclose}

\newcommand{\abs}[1]{\lvert#1\rvert}
\newcommand{\ket}[1]{\lvert#1\rangle}

\newcommand{\braket}[2]{\langle#1\mid#2\rangle}
\newcommand{\ketbra}[2]{\lvert#1\rangle\langle#2\rvert}

\newcommand{\expe}[1]{\langle#1\rangle}
\newcommand{\set}[1]{\{#1\}}
\DeclareMathOperator{\pr}{P}
\newcommand{\pf}{p}
\newcommand{\sect}{\S}
\newcommand{\sects}{\S\S}
\newcommand{\chap}{ch.}%
\newcommand{\eqn}{eq.}%
\newcommand{\eqns}{eqs.}%

\theoremstyle{remark}

\theoremstyle{definition}

\newcommand{\etc}{{etc.}}
\newcommand{\ie}{{i.e.}}
\newcommand{\Ie}{{I.e.}}
\newcommand{\Eg}{{E.g.}}
\newcommand{\eg}{{e.g.}}
\newcommand{\viz}{{viz.}}
\newcommand{\cf}{{cf.}}
\newcommand{\Cf}{{Cf.}}

\newcommand{\etal}{{et al.}}

\newcommand{\bd}{\hspace{0pt}}%

\providecommand{\href}[2]{#2}

\alleugreek

\renewcommand{\langlatin}{\foreignlanguage{nohyphenation}}
\usepackage[hypertex,breaklinks,pdfauthor={P G L  Porta Mana}]{hyperref}

\renewenvironment{acknowledgements}{\section*{Acknowledgements}}{\par}
\newcommand{\povm}{positive-\bd operator-\bd valued measure}
\newcommand{\povmm}{positive-\bd operator-\bd valued-\bd measure}

\newcommand{\so}{statistical operator}

\newcommand{\tsum}{{\textstyle\sum}}
\DeclareMathOperator*{\tprod}{\textstyle\prod}

\newcommand{\zrh}{\bm{\rho}}
\newcommand{\zrhh}{\rho}
\newcommand{\ze}{\bm{\varEpsilon}}
\newcommand{\zee}{\bm{\varEpsilon}}
\newcommand{\zeo}{\bm{\varEpsilon}}
\newcommand{\zen}{\bm{\varDelta}}
\newcommand{\zta}{\bm{\tau}}

\newcommand{\zone}{\ketbra{1}{1}}
\newcommand{\ztwo}{\ketbra{2}{2}}
\newcommand{\zthr}{\ketbra{3}{3}}

\newcommand{\zlgm}{\bm{\lambda}}
\newcommand{\zla}{x}
\newcommand{\zll}{\bm{x}}
\newcommand{\zllt}{\bm{x}'}
\newcommand{\zlat}{x'}
\newcommand{\zllc}{\hat{\bm{x}}}
\newcommand{\zxx}{\bm{y}}
\newcommand{\zA}{\bm{A}}
\newcommand{\zb}{\bm{b}}

\newcommand{\zlla}{\zll'}
\newcommand{\zllb}{\zll''}
\newcommand{\zrha}{\zrh'}
\newcommand{\zrhb}{\zrh''}
\newcommand{\zcoa}{\alpha'}
\newcommand{\zcob}{\alpha''}

\newcommand{\zhi}{d}
\newcommand{\zNv}{\bar{N}}

\newcommand{\zgi}{g_\text{co}}
\newcommand{\zgj}{g_\text{ga}}
\newcommand{\zI}{I_\text{co}}
\newcommand{\zJ}{I_\text{ga}}

\newcommand{\zsoset}{\mathbb{S}}
\newcommand{\zsosett}{{\mathbb{S}_3}}
\newcommand{\zxset}{{\mathbb{B}_8}}
\newcommand{\zcset}{{\mathbb{C}_8}}
\newcommand{\zsconv}{S}
\newcommand{\zrconv}{B}
\newcommand{\zchf}{\chi_{\mathbb{B}}}

\newcommand{\zve}{\omega}
\newcommand{\zm}{m}
\newcommand{\zciso}{c}
\newcommand{\zsubco}{B}

\newcommand{\origo}{\bm{O}}

\newcommand{\zpu}{\phi}
\newcommand{\zpa}{a}
\newcommand{\zpb}{b}
\newcommand{\zpc}{c}
\newcommand{\zfb}{\beta}
\newcommand{\zfc}{\gamma}

\newcommand{\zP}{\bm{v}}

\newcommand{\zq}{q}
\newcommand{\zqq}{\bm{\zq}}
\newcommand{\zf}{\bm{f}}
\newcommand{\zfi}{\bm{f}^*}
\newcommand{\zffi}{f^*}
\newcommand{\zzf}{\varPhi}
\newcommand{\zzfi}{{\varPhi_\infty}}

\newcommand{\zux}{\epsilon_3}
\newcommand{\zuy}{\epsilon_8}

\begin{document}
\bibliographystyle{apsrevmananum} 

\title{Numerical Bayesian state assignment for a three-level quantum system
  \\ I. Absolute-frequency data; constant and Gaussian-like priors}

\author{\firstname{A.} \surname{M{\aa}nsson}}
\email{andman@imit.kth.se}

\author{\firstname{P. G. L.} \surname{Porta Mana}} 
\email{mana@kth.se}

\author{\firstname{G.} \surname{Bj\"{o}rk}}

\affiliation{Kungliga Tekniska H\"ogskolan, Isafjordsgatan
  22, SE-164\,40 Stockholm, Sweden}

\date{29 April 2007}

\begin{abstract}
  This paper offers examples of concrete numerical
  applications of Bayesian quantum-state-assignment
  methods to a three-level quantum system. The \so\
  assigned on the evidence of various measurement data and
  kinds of prior knowledge is computed partly
  analytically, partly through numerical integration (in
  eight dimensions) on a computer. The measurement data
  consist in absolute frequencies of the outcomes of $N$
  identical von~Neumann projective measurements performed
  on $N$ identically prepared three-level systems. Various
  small values of $N$ as well as the large-$N$ limit are
  considered. Two kinds of prior knowledge are used:
  one represented by a plausibility distribution constant
  in respect of the convex structure of the set of \so s;
  the other represented by a Gaussian-like distribution
  centred on a pure \so, and thus reflecting a situation
  in which one has useful prior knowledge about the likely
  preparation of the system.

In a companion paper the case of
measurement data consisting in average values, and an
additional prior studied by Slater, are considered.
\end{abstract}

\pacs{03.67.-a,02.50.Cw,02.50.Tt,05.30.-d,02.60.-x}

\maketitle

\section{Introduction}
\label{sec:intro}

\subsection{Quantum-state assignment: theory\ldots}
\label{sec:qsa-th}

A number of different ``quantum-state assignment'' (or
``reconstruction'', ``estimation'', ``retrodiction'')
techniques have been studied in the literature. Their
purpose is to \emph{encode} various kinds of measurement
data and prior knowledge, especially in cases in which the
former is meager, into a statistical operator (or
``density matrix'') suitable for deriving the
plausibilities of future or past measurement outcomes. The
use of probabilistic methods is clearly essential in this
task,\footnote{``Quantum-state tomography'' (\cf\
  \eg~\citep{leonhardt1997,jamesetal2001}) usually refers
  to the special case in which, roughly speaking, the
  number of measurements and measurement outcomes are
  sufficient to yield a unique statistical operator.
  Mathematically, we have a well-posed inverse problem
  that does not require plausible reasoning. This case is
  only achieved as the number of outcomes gets larger and
  larger.} and they are implemented in a variety of ways.
There are implementations based on
maximum-relative-entropy methods\footnote{The literature
  on these is so vast as to render any small sample very
  unfair. Early and latest contributions
  are~\citep{jaynes1957b,jaynes1980c,derkaetal1996,buzeketal1997,buzeketal1998,buzeketal2000}.}
and others based on more general Bayesian
methods~\citep{jeffreys1931_r1957,jeffreys1939_r1998,jaynes1994_r2003,definetti1970_t1990,bernardoetal1994,gelmanetal1995_r2004,gregory2005}.
Here we are concerned with the latter, which can
apparently be used with a larger variety of prior
knowledge than the former.\footnote{\Eg, for a spin-1/2
  system, knowledge that ``the state that holds is either
  the one represented by (the \so) $\ketbra{z^+}{z^+}$ or
  the one represented by $\ketbra{z^-}{z^-}$'', is
  different from knowledge that ``the state that holds is
  either the one represented by $\ketbra{x^+}{x^+}$ or the
  one represented by $\ketbra{x^-}{x^-}$'', and this
  difference can be usefully exploited in some situations:
  Make a measurement corresponding to the \povm\
  $\set{\ketbra{z^+}{z^+}, \ketbra{z^-}{z^-}}$, and
  suppose you obtain the `$z^+$' result. Conditional on the
  first kind of prior knowledge you then know that ``the
  original state was the one represented by
  $\ketbra{z^+}{z^+}$'', whereas conditional on the second
  you know now just as much as before. But in quantum
  maximum-entropy methods both kinds of prior knowledge
  are encoded in the same way, \viz\ as the same
  ``completely mixed'' \so\ to be used with the quantum
  relative entropy; these methods thus provide less
  predictive power in this example.} (Old statistical
methods, like maximum likelihood, are not considered here
either since they are only special cases of the Bayesian
ones.)

The fundamental ideas behind the Bayesian techniques were
developed gradually. A sample of more or less related
studies could consist in the works by
Segal~\citep{segal1947},
Helstrom~\citep{helstrom1967,helstrom1974,helstrom1976},
Band and
Park~\citep{parketal1971,bandetal1971,bandetal1976,parketal1976,parketal1977,bandetal1977,bandetal1979,parketal1980},
Holevo~\citep{holevo1973b,holevo1980_t1982,holevo2001},
Bloore~\citep{bloore1976},
Ivanovi\'c~\citep{ivanovic1981,ivanovic1983,ivanovic1984,ivanovic1987},
Larson and Dukes~\citep{larsonetal1991},
Jones~\citep{jones1991b,jones1994}, Malley and
Hornstein~\citep{malleyetal1993},
Slater~\citep{slater1993,slater1995b}, and many
others~\citep{mackey1963,mielnik1968,mielnik1974,davies1978,harriman1978,harriman1978b,harriman1979,harriman1983,harriman1984,balianetal1987b,balian1989c,derkaetal1996,buzeketal1997,buzeketal1998,derkaetal1998,buzeketal1999,barnettetal2000,barnettetal2000b,buzeketal2000,harriman2001,schacketal2001,cavesetal2002,peggetal2002b,vanenketal2002,vanenketal2002b,man'koetal2004,tanakaetal2005b,man'koetal2006};
some central points can already be found in
Bloch~\citep{bloch1989_r2000}. Such a dull list
unfortunately does not do justice to the relative
importance of the individual contributions (some of which
are just rediscoveries of earlier ones); those by
Helstrom, Holevo, Larson and Dukes, and Jones, however,
deserve special mention.

All Bayesian quantum-state assignment techniques more or
less agree in the expression used to calculate the \so\
$\zrh_{D\land I}$ encoding the measurement data $D$ and
the prior knowledge $I$. The `conditions', or `states', in
which the system can be prepared are represented by \so s
$\zrh$, whose set we denote by $\zsoset$. Let the prior
knowledge $I$ about the possible state in which the system
is prepared be expressed by a `prior' plausibility
distribution $\pf(\zrh \cond I)\, \di\zrh = g(\zrh)\,
\di\zrh$ (where $\di\zrh$ is a volume element on $\zsoset$
or a subset thereof, and $g$ a plausibility density; more
technical details are given in \sect~\ref{sec:prior}). Let
the measurement data $D$ consist in a set of $N$ outcomes
$i_1, \dotsc, i_k, \dotsc, i_N$ of $N$ measurements,
represented by the $N$ \povm s $\set{\ze^{(k)}_\mu \with
  \mu = 1, \dotsc, r_k}$, $k= 1, \dotsc, N$. Bayesian
quantum-state assignment techniques yield a `posterior'
plausibility distribution of the form
\begin{subequations}\label{eq:yield}
\begin{equation}\label{eq:yield_post_pdf}
  \begin{split}
    \pf(\zrh \cond D \land I)\,\di\zrh 
&=
 \frac{ \pf(D
      \cond \zrh)\, \pf(\zrh \cond I)\, \di\zrh } {\int
      \pf(D \cond \zrh)\, \pf(\zrh \cond I)\, \di\zrh},
\\
&=
\frac{
\Bigl[\tprod_k \tr\bigl(\ze^{(k)}_{i_k} \zrh
      \bigr)\Bigr]\, g(\zrh)\, \di\zrh 
}
{
\int \Bigl[\tprod_k
      \tr\bigl(\ze^{(k)}_{i_k} \zrh \bigr)\Bigr]\,
      g(\zrh)\, \di\zrh}.
    \end{split}
\end{equation}
  and a \so\ $\zrh_{D\land I}$ given by a sort of weighted
  average,\footnote{Note that, as shown in
    \sect~\ref{sec:scenario}, the derivation of the
    formula for $\zrh_{D\land I}$ does not require
    decision-theoretical concepts.}
\begin{equation}
  \label{eq:yield_SO}
\zrh_{D \land I}
\defd
\int \zrh\,\pf(\zrh \cond D \land I)\,\di\zrh 
=
\frac{\int \zrh
\Bigl[\tprod_k \tr\bigl(\ze^{(k)}_{i_k} \zrh
      \bigr)\Bigr]\, g(\zrh)\, \di\zrh 
}
{
\int \Bigl[\tprod_k
      \tr\bigl(\ze^{(k)}_{i_k} \zrh \bigr)\Bigr]\,
      g(\zrh)\, \di\zrh}.
\end{equation}
\end{subequations}
These formulae may present differences of detail from author
to author, reflecting --- quite excitingly!\ --- different
philosophical stands. For instance, the prior distribution
(and therefore the integration) is in general defined over
the whole set of statistical operators; but a person who
conceives only \emph{pure} statistical operators as
representing sort of ``real, internal (microscopic) states
of the system'' may restrict it to those only. A person
who, on the other hand, thinks of the statistical
operators themselves as encoding ``states of knowledge''
\french{au pair} with plausibility distributions, might
see the prior distribution as a ``plausibility of a
plausibility'', and thus prefer to derive the formula
above through a quantum analogue of de~Finetti's theorem;
in this case the derivation will involve a tensor product
$\zrh \otimes \dotsb \otimes \zrh$ of multiple copies of
the same statistical operator.\footnote{We leave to the
  reader the entertaining task of identifying these
  various philosophical stances in the references already
  provided.}

The formulae~\eqref{eq:yield} (or special cases thereof) are proposed and
used in Larson and Dukes~\citep{larsonetal1991},
Jones~\citep{jones1991b,jones1994}, and \eg\ Slater~\citep{slater1995b},
Derka, Buzek, \etal~\citep{derkaetal1996,buzeketal1998,buzeketal1999}, and
Mana~\citep{mana2004b}. We arrived at these same formulae (as special cases
of formulae applicable to generic, not necessarily quantum-theoretical
systems) in a series of
papers~\citep{portamanaetal2006,portamanaetal2007,portamana2007b,portamana2007}
(see also~\citep{mana2003,mana2004b}) in which we studied and tried to
solve the various philosophical issues to \emph{our} satisfaction.

\subsection{\ldots and practice}
\label{sec:practice}

In regard to the numerical computation of
formulae~\eqref{eq:yield} in actual or fictive
state-assignment problems, with explicitly given prior
distributions and measurement data, the number of studies
is much smaller. The main problem is that
formula~\eqref{eq:yield_SO}, when applied to a
$\zhi$-level system, generally involves an integration
over a complicated (see \eg\ figs.~\ref{fig:sections} and
those
in~\citep{jakobczyketal2001,kimura2003,kimuraetal2004,portamana2006b})
convex region of $\zhi^2-1$ dimensions ($2\zhi -2$
dimensions if only pure \so s are considered), and one has
to choose between explicit integration limits but very
complex integrands, or, \latin{vice versa}, simpler
integrands but implicitly defined integration regions.

Therefore explicit calculations have hitherto been
confined almost exclusively to two-level systems, which
have the obvious advantages of low-dimensionality and
symmetry (the set of \so s is the three-dimensional Bloch
ball~\citep{bloch1946,blochetal1946}); in some cases these allow the
derivation of analytical results.\footnote{The high
  symmetry, however, renders the results independent of
  the particular choice of prior knowledge in some cases,
  \eg\ when the prior is spherically symmetric and the
  data consist on averages.} In studies by
Jones~\citep{jones1991b}, Larson and
Dukes~\citep{larsonetal1991},
Slater~\citep{slater1995b,slater1996,slater1996b} (these
are very interesting studies; \cf\
also~\citep{slater1997,slater1997b}), and Bu\v{z}ek,
Derka,
\etal~\citep{derkaetal1996,buzeketal1998,buzeketal1999},
the posterior distributions $\pf(\zrh \cond D)\, \di\zrh$
and the ensuing \so\ $\zrh_{D \land I}$
are explicitly calculated for measurement data $D$ and
priors $I$ of various kinds. In some of these studies the
integrations range over the whole set of \so s, in others over
the pure ones only. As regards higher-level systems, the
only numerical study known to us is that by Bu\v{z}ek
\etal~\citep{buzeketal1998,buzeketal1999} for a spin-3/2
system; however, they assume from the start that the
\latin{a priori} possible \so s are confined to a
three-dimensional subset of the pure ones; this assumption
simplifies the integration problem from 15 to 3
dimensions.


\subsection{More practice: the place of the present study}
\label{sec:pres_study}

\begin{figure*}[pht!]
  \includegraphics*[height=0.31\textheight, trim= 1em 5em 1em
  5em]{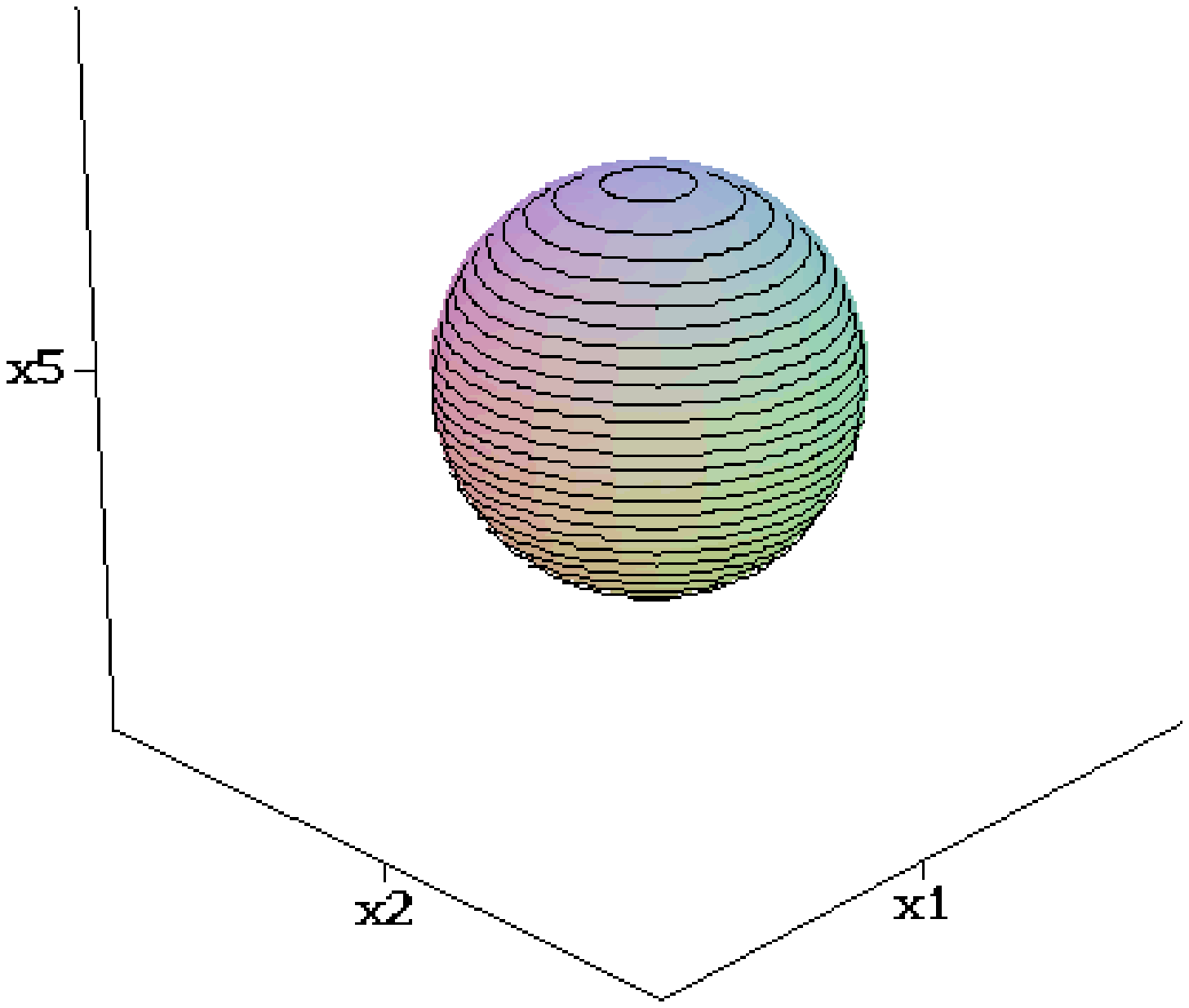}
  \includegraphics*[height=0.31\textheight, trim= 1em 5em 1em
  5em]{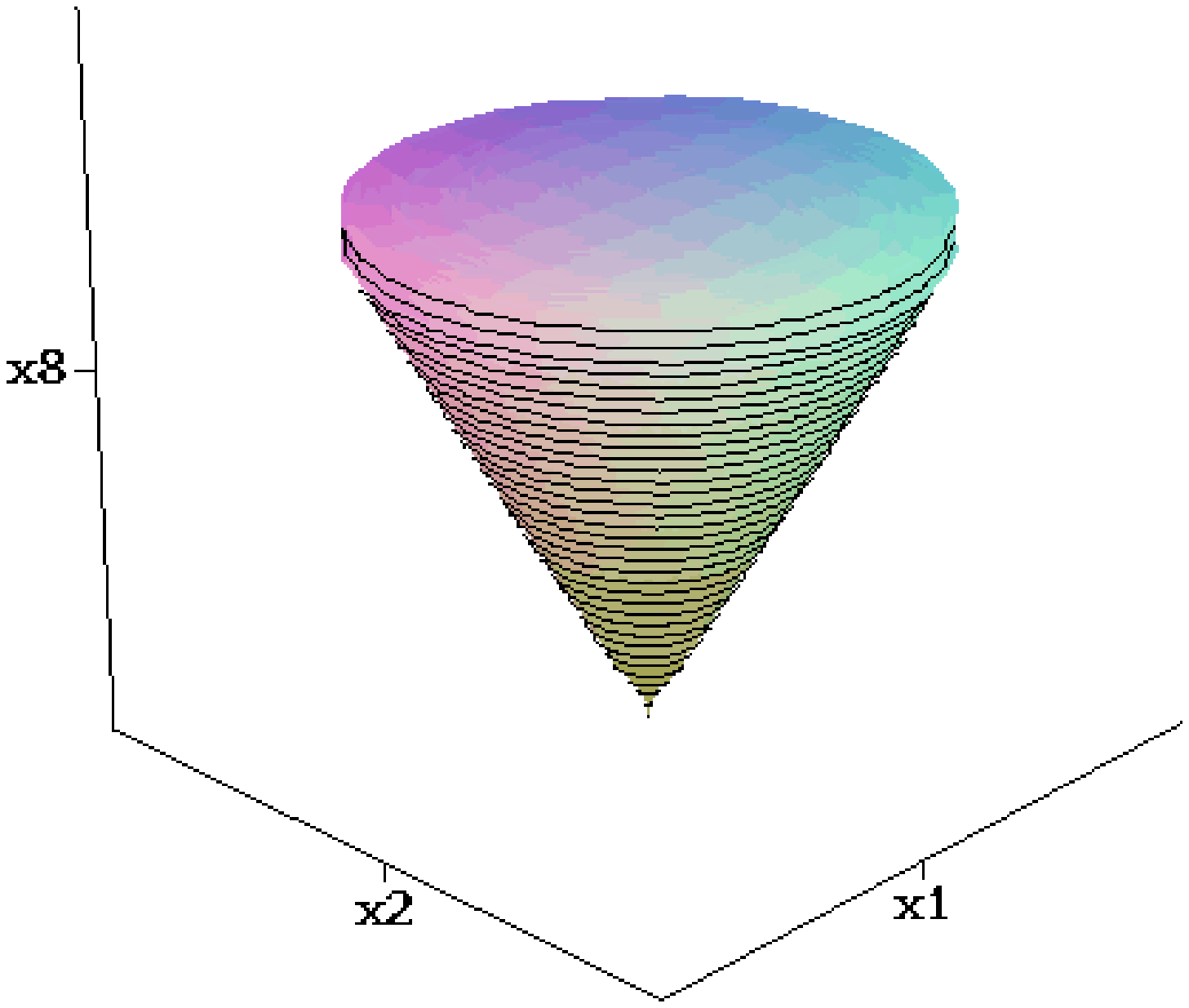}
\\
  \includegraphics*[height=0.31\textheight, trim= 1em 5em 1em
  5em]{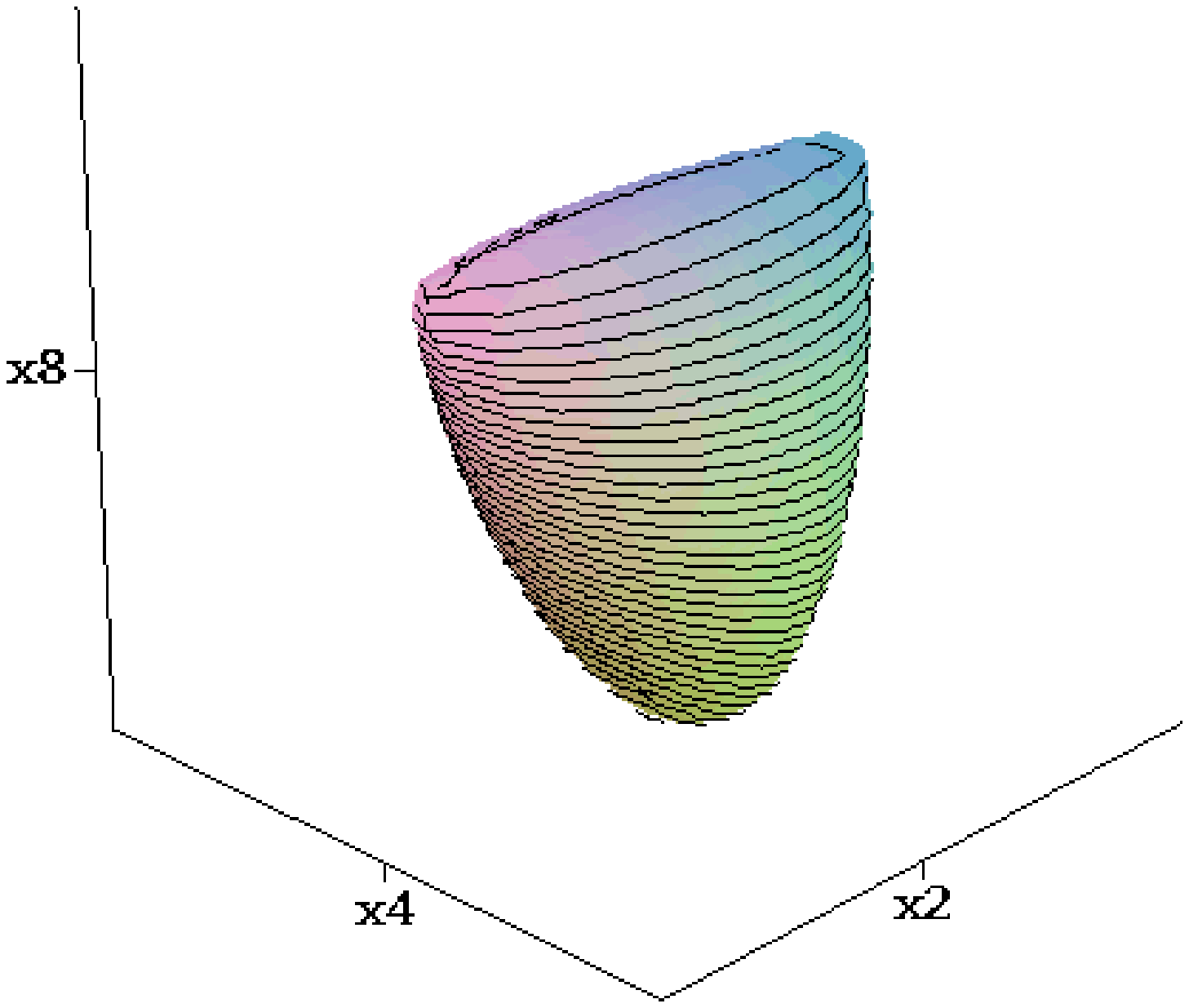}
  \includegraphics*[height=0.31\textheight, trim= 1em 5em 1em
  5em]{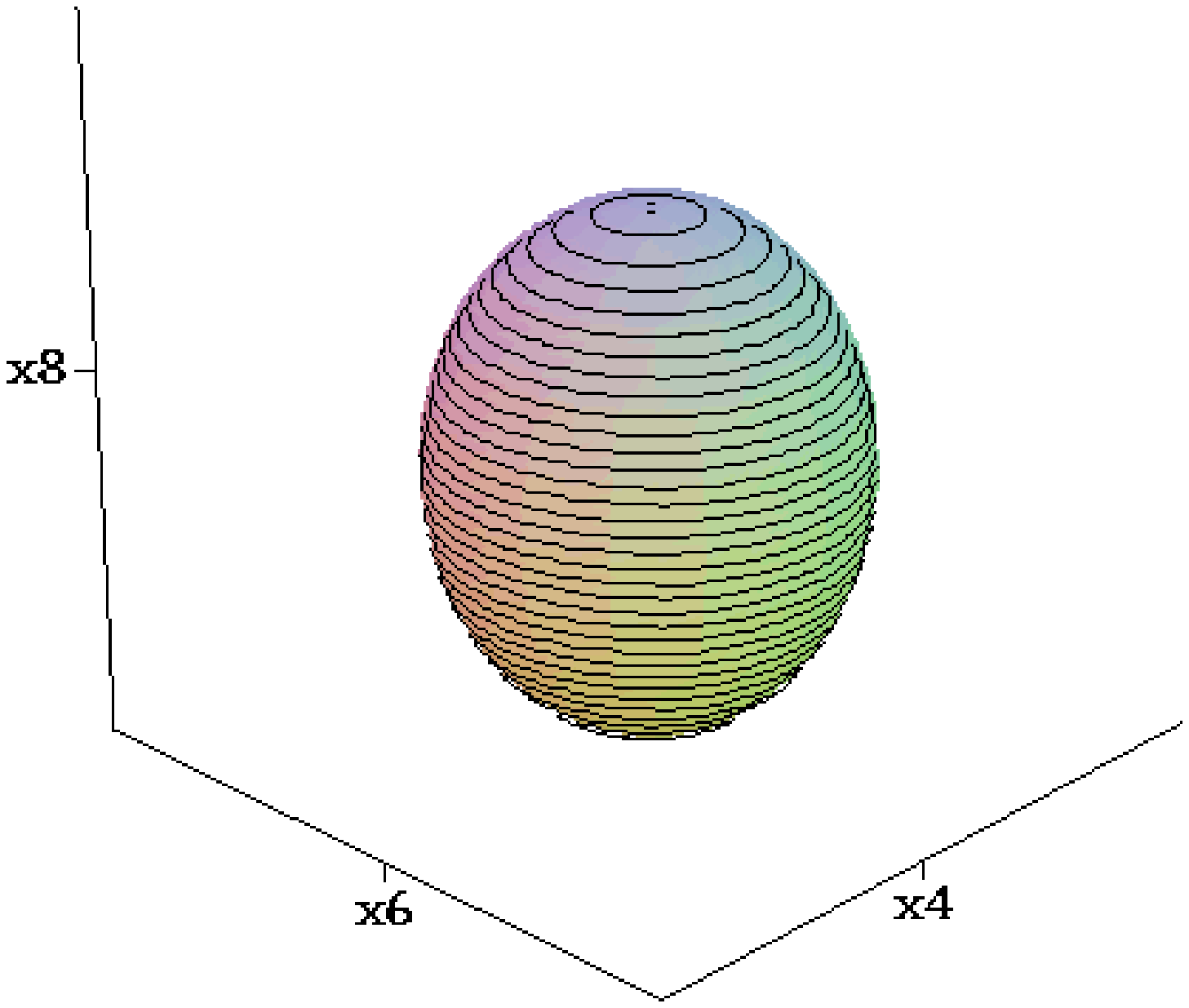}
\\
  \includegraphics*[height=0.31\textheight, trim= 1em 5em 1em
  5em]{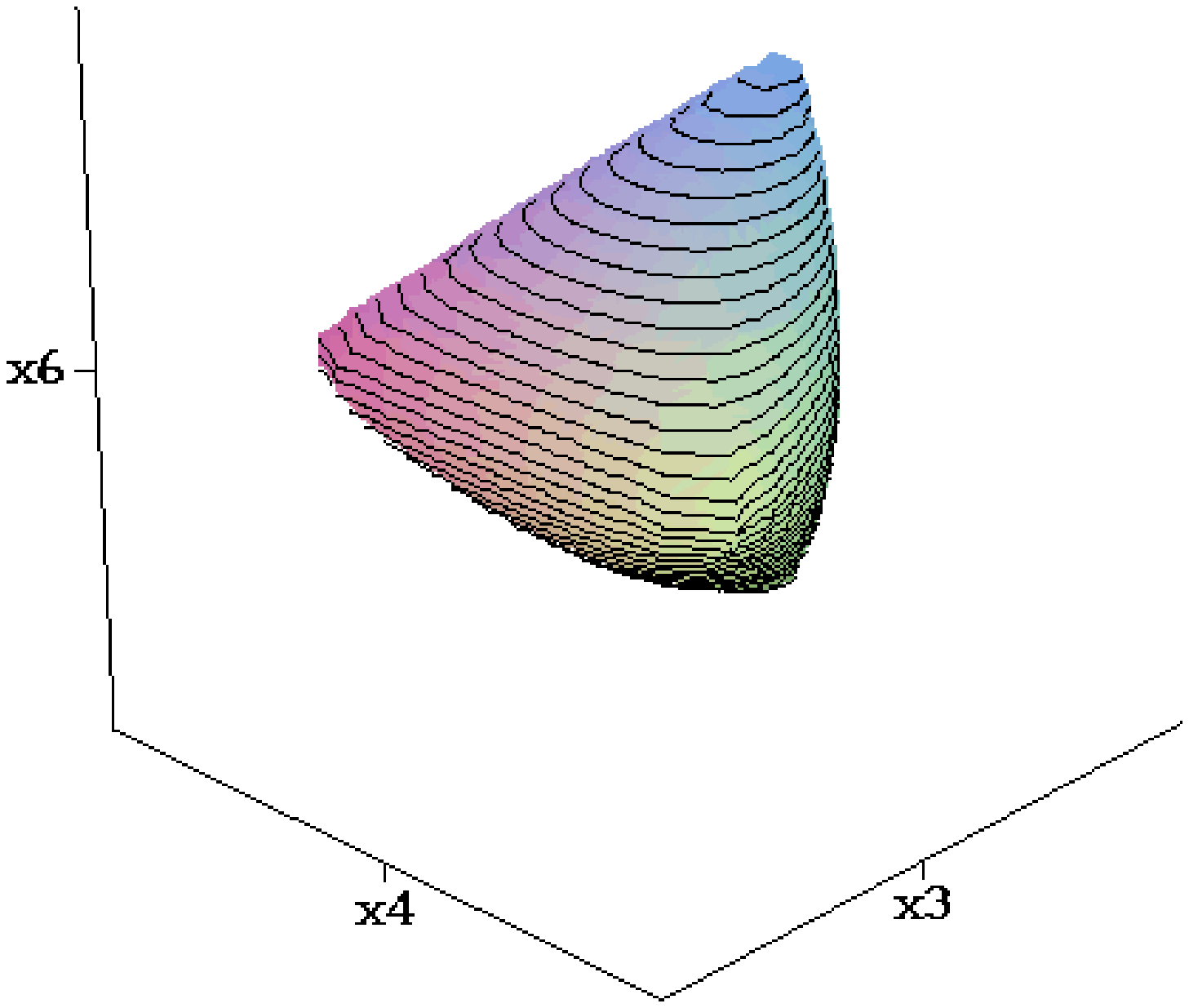}
  \includegraphics*[height=0.31\textheight, trim= 1em 5em 1em
  5em]{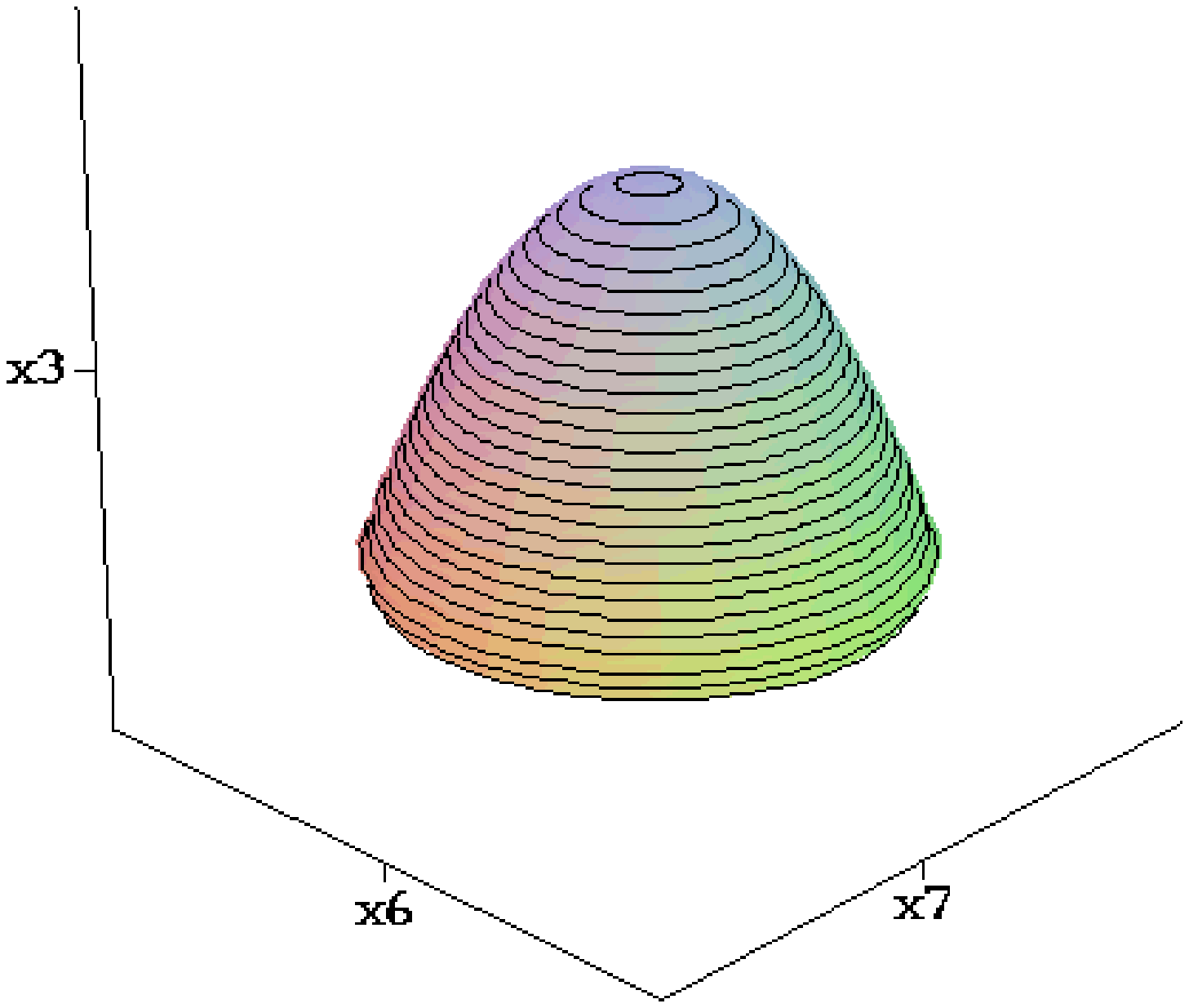}
  \caption{Some three-dimensional sections of the
    eight-dimensional set $\zsosett$ of the \so s for a
    three-level quantum system. The adopted coordinate system is
    explained in \sect~\ref{sec:blochvectors}.}
  \label{fig:sections}
\end{figure*}

In this paper and its companion~\citep{maanssonetal2007} we provide
numerical examples of quantum-state assignment, via \eqns~\eqref{eq:yield},
for a three-level system. The set of \so s of such a system, $\zsosett$, is
eight-dimensional --- a high but still computationally tractable number of
dimensions --- and has less symmetries, in respect of its dimensionality,
than that of a two-level one: a two-level system is a ball in $\RR^3$, but
a three-level one is definitely not a ball in $\RR^8$.\footnote{In
  group-theoretical terms, the ``quantum'' symmetries of the set of \so s
  of a three-level system are fewer than those it could have had as a
  eight-dimensional compact convex set. The former symmetries are in fact
  equivalent to the group
  $\mathrm{U}(3)/\mathrm{U}(1)$~\citep{bengtsson_c2006,wigner1931_t1959,kadison1965,hunziker1972}\citep[\sect~4]{baez2002},
  of 8 dimensions, whereas the latter could have been as large the group
  $\mathrm{SO}(8)$, of 28
  dimensions~\citep{gilmore1974,curtis1979_r1984,baker2002,hall2003_r2004}.
  Compare with the case of a two-level system, whose symmetry group
  $\mathrm{U}(2)/\mathrm{U}(1)$, of 3 dimensions, is isomorphic to the
  largest symmetry group that a three-dimensional compact convex body can
  have, $\mathrm{SO}(3)$. (We have only considered the connected part of
  these groups; one should also take the semidirect product with $\ZZ_2$.)}
Some three-dimensional sections of this eight-dimensional set are given in
fig.~\ref{fig:sections} (two-dimensional sections can be found
in~\citep{kimura2003}; four-dimensional ones are also
available~\citep{portamana2006b}); see also Bloore's very interesting
study~\citep{bloore1976}.

We study data $D$ and prior knowledge $I$ of the following
kind:
\begin{itemize}
\item The measurement data $D$ consist in a set of $N$
  outcomes of $N$ instances of the same measurement
  performed on $N$ identically prepared systems. The
  measurement is represented by the extreme \povm\ (\ie,
  non-degenerate `von~Neumann measurement') having three
  possible distinct outcomes $\set{\text{`1'}, \text{`2'},
    \text{`3'}}$ represented by the eigenprojectors
  $\set{\ketbra{1}{1}, \ketbra{2}{2}, \ketbra{3}{3}}$. The
  data $D$ thus correspond to a triple of absolute
  frequencies $(N_1, N_2, N_3) \defs \zNv$, with $N_i \ge
  0$ and $\sum_i N_i = N$. We consider various such
  triples 
  for small values of $N$, as well as for the limiting case
  of very large $N$.
\item Two different kinds of prior knowledge $I$ are used.
  The first, $\zI$, is represented by a prior plausibility
  distribution
\begin{equation}
  \label{eq:first_prior_rho}
  \pf(\zrh \cond \zI)\, \di\zrh 
= \zgi(\zrh)\, \di\zrh
\propto \di\zrh,
\end{equation} 
which is constant in respect of the convex structure of
the set of statistical operators, in the sense explained
in \sects~\ref{sec:blochvectors} and~\ref{sec:prior}.
  The second, $\zJ$, is represented by a spherically symmetric,
  Gaussian-like prior distribution
\begin{equation}\label{eq:second_prior_rho}
    \pf(\zrh \cond \zJ)\, \di\zrh
= \zgj(\zrh)\, \di\zrh
\propto
    \exp\biggl\{
-\frac{\tr[(\zrh -\ztwo)^2]}{s^2}
\biggr\}\, \di\zrh,
\end{equation}
centred on the statistical operator $\ztwo$, one of the
projectors of the von~Neumann measurement. This prior
expresses some kind of knowledge that leads us to assign
higher plausibility to regions in the vicinity of $\ztwo$.
\end{itemize}


To assign a \so\ $\zrh_{D \land I}$ from these data and
priors means to assign eight independent real coefficients
of its matrix elements, or equivalently a vector of eight
real parameters bijectively associated with them. These
parameters, according to \eqn~\eqref{eq:yield_SO}, must be
computed by the integration of a function (actually two,
the other being a normalisation factor) defined over the
set of all \so s. Hence the function itself and the
integration region can be expressed in terms of eight
coordinates, corresponding to the parameters. The
coordinate system should be chosen in such a way that both
the function and the integration limits have a not too
complex form. For these reasons we choose the
parametrisation studied in particular by
Kimura~\citep{kimura2003}. In this case the vectors of
real parameters associated to a \so\ is called a `Bloch
vector'.

In such a coordinate system, six of the eight parameters
can be calculated analytically and quite straightforwardly
by symmetry arguments, for all absolute-frequency triples
$\zNv$. The remaining two parameters have been numerically
calculated for some triples $\zNv$ by a computer using
quasi-Monte~Carlo integration methods, suitable for
high-dimensional problems. Further symmetry arguments
yield the parameters for the remaining triples.


All these points as well as the results are discussed in
the paper as follows: In \sect~\ref{sec:scenario} we
quickly present the reasoning leading to the
statistical-operator-assignment formulae~\eqref{eq:yield},
and particularise the latter to our study. In
\sect~\ref{sec:blochvectors} Kimura's parametrisation and
the Bloch-vector set are introduced. The two prior
distributions adopted are discussed in
\sect~\ref{sec:prior}. The calculation, by symmetry
arguments and by numerical integration, of the Bloch
vectors and of the corresponding \so s is presented in
\sect~\ref{sec:computation}, for all data and priors. In
\sect~\ref{sec:measerr} we offer some remarks on the
incorporation into the formalism of uncertainties in the
detection of outcomes. In \sect~\ref{sec:N_infinity} we
discuss the form the assigned \so\ takes in the limit of a
very large number of measurements. Finally, the last
section summarises and discusses the main points and
results.


\section{Statistical-operator assignment\label{sec:scenario}}

\subsection{General case}
\label{sec:scen_gen_case}

This section provides a summary derivation of the formulae for
statistical-operator assignment. For a more general derivation of analogous
formulae valid for any kind of system (classical, quantum, or exotic), and
for a discussion of some philosophical points involved, we refer the reader
to~\citep{portamanaetal2006,portamanaetal2007,portamana2007b,portamana2007}
and also~\citep{mana2003,mana2004b}.

There is a preparation scheme that produces quantum
systems always in the same `condition' --- the same
`state'. We do not know which this condition is, amongst a
set of possible ones;\footnote{We intentionally use the
  vague term `condition', since each researcher can
  understand it in terms of his or her favourite physical
  picture (internal microscopic configurations,
  macroscopic procedures, pilot waves, propensities,
  grounds for judgements of exchangeability, or whatnot).
  Quantum theory offers no concrete physical picture,
  only some constraints on how such a picture should work;
  so each one can provide one's favourite.} although there
may be some conditions in that set that are more plausible
than others. Our knowledge $I$, in other words, is
expressed by a plausibility distribution over these
conditions. To each condition is associated a \so; this
encodes the plausibility distributions that we assign for
all possible quantum measurements, given that that
particular condition hold. Therefore we can and shall more
simply speak in terms of \so s instead of the respective
conditions. Note that this is, however, a metonymy, \ie\
we are speaking about something (`\so') although it is
something else but related to it (`condition') that we
really mean.

We thus have a plausibility distribution over some \so s.
It can in full generality be written as
\begin{equation}
  \label{eq:plaus_prior_so}
  \pf(\zrh \cond I)\, \di\zrh
= g(\zrh)\, \di\zrh,
\end{equation}
defined over the \emph{whole} set of \so s, denoted by
$\zsoset$. The function $g$ is a normalised positive
generalised function.\footnote{See
  footnotes~\ref{fn:egorov} and~\ref{fn:soffer}.} In this
way the more general case is also accounted for in which
the whole set of \so s $\zsoset$ is involved: the case
with a finite number of \latin{a priori} possible \so s
corresponds to a $g$ equal to a sum of appropriately
weighted Dirac deltas.%
\footnote{\label{fn:mult_so}The knowledge $I$ and all
  inferential steps to follow concern a preparation scheme
  in general and not specifically this or that system
  only; just like tastings of cakes made according to a
  given unknown recipe increase our knowledge of the
  recipe, not only of the cakes. If one insists in seeing
  the knowledge $I$ and the various inferences as
  referring to a given set of, say, $M$ systems only, then
  that knowledge is represented by a plausibility
  distribution over the \so s of these $M$ systems, \ie\
  over the Cartesian product $\zsoset^M$, and has the form
  $\pf(\zrh^{(1)}, \dotsc, \zrh^{(M)} \cond
  I)\,\di\zrh^{(1)} \dotsm \di\zrh^{(M)}= g(\zrh^{(1)})\,
  \delt(\zrh^{(2)} - \zrh^{(1)}) \dotsm \delt(\zrh^{(M)} -
  \zrh^{(1)}) \,\di\zrh^{(1)} \dotsm \di\zrh^{(M)}$.
  Integrations are then also to be understood accordingly.
  Note moreover that if we consider joint quantum
  measurements on all the systems together, then we are
  really dealing with \emph{one} quantum system, not $M$.}

Our `prior' knowledge $I$ about the preparation can be
represented by a unique \so: Suppose we are to give the
plausibility of the $\mu$th outcome of an arbitrary
measurement, represented by the \povm\ $\set{\zeo_{\mu}}$,
performed on a system produced according to the
preparation. Quantum mechanics dictates the plausibilities
$\pf(\zeo_\mu \cond \zrh) = \tr(\zeo_\mu \zrh)$, and by
the rules of plausibility theory we assign, conditional on
$I$,\footnote{We do not explicitly write the prior
  knowledge $I$ whenever the \so\ appears on the
  conditional side of the plausibility; \ie, $\pf(\cdot
  \cond \zrh) \defd \pf(\cdot \cond \zrh, I)$.}
\begin{equation}
  \label{eq:pred_gen_meas}
  \pf(\zeo_\mu \cond I) =
\int_{\zsoset} \pf(\zeo_\mu \cond \zrh)\,\pf(\zrh \cond I)\,
\di\zrh
=
\int_{\zsoset} \tr(\zeo_\mu \zrh)\, g(\zrh)\, \di\zrh,
\end{equation}
or more compactly, by linearity of the trace,
\begin{gather}
  \label{eq:more_comp}
  \begin{split}
    \pf(\zeo_\mu \cond I) 
&=\tr\bigl[\zeo_\mu\smallint \zrh\, g(\zrh)\, \di\zrh \bigr],\\
&= \tr(\zeo_\mu \zrh_I),
  \end{split}
\\
\intertext{with the \so\ $\zrh_I$ defined as}
\zrh_I
\defd
\int_{\zsoset} \zrh\, \pf(\zrh \cond I)\, \di\zrh
=
\int_{\zsoset} \zrh\, g(\zrh)\, \di\zrh.
\label{eq:def_rho_I}
\end{gather}
The prior knowledge $I$ can thus be compactly represented
by, or ``encoded in'', the \so\ $\zrh_I$. Note how
$\zrh_I$ appears naturally, without the need to invoke
decision-theoretics arguments and concepts, like cost
functions \etc{} Note also that the association between
$I$ and $\zrh_I$ is by construction valid for generic
knowledge $I$, be it ``prior'' or not.

The \so\ $\zrh_I$ is a ``disposable'' object. As soon as
we know the outcome of a measurement on a system produced
according to our preparation, the plausibility
distribution $\pf(\zrh \cond I)\, \di\zrh$ should be
updated on the evidence of this new piece of data $D$,
and thus we get a new \so\ $\zrh_{I\land D}$. And so on.
It is a fundamental characteristic of plausibility theory
that this update can indifferently be performed with a
piece of data at a time or all at once.

So suppose we come to know that $N$ measurements,
represented by the $N$ \povm s $\set{\ze^{(k)}_\mu \with
  \mu = 1, \dotsc, r_k}$, $k= 1, \dotsc, N$, are or have
been performed on $N$ systems for which our knowledge $I$
holds. Note that some, even all, of the measurements (and
therefore their \povm s) can be identical. The outcomes
$i_1, \dotsc, i_k, \dotsc, i_N$ are or were obtained; this
is our new data $D$. The plausibility for this to occur,
according to the prior knowledge $I$, is given by a
generalisation of expression~\eqref{eq:pred_gen_meas}:
\begin{gather}
  \label{eq:many_meas}
\pf(D \cond I) \equiv   \pf\bigl(\ze^{(1)}_{i_1}, \dotsc, \ze^{(N)}_{i_N} \cond
   I\bigr) = 
\int_{\zsoset} 
\Bigl[\tprod_{k=1}^N \pf\bigl(\ze^{(k)}_{i_k} \cond
\zrh\bigr)\Bigr]\,
 \pf(\zrh \cond I)\,
\di\zrh.
\end{gather}
On the evidence of $D$ we can update the prior plausibility
distribution $\pf(\zrh \cond I)\,
\di\zrh$. By the rules of plausibility theory
\begin{equation}\label{eq:update_prior}
  \begin{split}
    \pf(\zrh \cond D \land I)\,\di\zrh 
&=
 \frac{ \pf(D
      \cond \zrh)\, \pf(\zrh \cond I)\, \di\zrh } {\int_{\zsoset}
      \pf(D \cond \zrh)\, \pf(\zrh \cond I)\, \di\zrh},
\\
&=
\frac{
\Bigl[\tprod_k \tr\bigl(\ze^{(k)}_{i_k} \zrh
      \bigr)\Bigr]\, g(\zrh)\, \di\zrh 
}
{
\int_{\zsoset} \Bigl[\tprod_k
      \tr\bigl(\ze^{(k)}_{i_k} \zrh \bigr)\Bigr]\,
      g(\zrh)\, \di\zrh}.
    \end{split}
\end{equation}

The \so\ encoding the joint knowledge $D \land I$ is thus,
according to \eqn~\eqref{eq:def_rho_I} and using
\eqn~\eqref{eq:update_prior},
\begin{equation}
  \label{eq:rho_D}
\zrh_{D \land I}
\defd
\int_{\zsoset} \zrh\,\pf(\zrh \cond D \land I)\,\di\zrh 
=
\frac{\int_{\zsoset} \zrh
\Bigl[\tprod_k \tr\bigl(\ze^{(k)}_{i_k} \zrh
      \bigr)\Bigr]\, g(\zrh)\, \di\zrh 
}
{
\int_{\zsoset} \Bigl[\tprod_k
      \tr\bigl(\ze^{(k)}_{i_k} \zrh \bigr)\Bigr]\,
      g(\zrh)\, \di\zrh}.
\end{equation}

\subsection{Three-level case}
\label{sec:3-lev_case}

So far everything has been quite general. Let us now
consider the particular cases studied in this paper.

The preparation scheme concerns three-level quantum
systems; the corresponding set of \so s will be denoted by
$\zsosett$. The $N$ measurements considered here are all
instances of the same measurement, namely a non-degenerate
projection-valued measurement (often called `von~Neumann
measurement'). Thus, for all $k= 1, \dotsc, N$,
$\set{\ze^{(k)}_\mu} = \set{\ze_\mu} \defd \set{\zone,
  \ztwo, \zthr}$. The projectors $\zone$, $\ztwo$, $\zthr$
define an orthonormal basis in Hilbert space. All relevant
operators will, quite naturally and advantageously, be
expressed in this basis. We have for example that
$\tr(\ze_{\mu} \zrh) \equiv \zrhh_{\mu \mu}$, the $\mu$th
diagonal element of $\zrh$.

The data $D$ consist in the set of outcomes
$\set{i_1, \dotsc, i_N}$ of the $N$ measurements, where
each $i_k$ is one of the three possible outcomes `1', `2',
or `3'. The formula~\eqref{eq:rho_D} for the  \so\ thus
takes the form
\begin{equation}
  \label{eq:rho_D_S_particular}
\zrh_{D \land I}
=
\frac{\int_{\zsosett} \zrh\,
\Bigl[
\tprod_{k=1}^N \zrhh_{i_k i_k}
\Bigr]\,
g(\zrh)\, \di\zrh 
}
{
\int_{\zsosett} \Bigl[\tprod_{k=1}^N \zrhh_{i_k i_k}
\Bigr]\,
      g(\zrh)\, \di\zrh},
\end{equation}
with $i_k \in \set{1,2,3}$ for all $k$.

However, it is clear from the expressions in the integrals
above that the exact order of the sequence of `1's, `2's,
and `3's is unimportant; only the absolute frequencies
$(N_1, N_2, N_3)$ of appearance of these three possible outcomes
matter (naturally, $N_i \ge 0$ and $\tsum_i N_i = N$). We
can thus rewrite the last equation as
\begin{equation}
  \label{eq:rho_D_S_particular_exp}
\zrh_{D \land I}
=
\frac{
\int_\zsosett
 \zrh\,
\Bigl[\tprod_{i=1}^3 \zrhh_{i i}^{N_i}\Bigr]\,
g(\zrh)\, \di\zrh
}{
\int_\zsosett
\Bigl[\tprod_{i=1}^3 \zrhh_{i i}^{N_i}\Bigr]\,
g(\zrh)\, \di\zrh
},
\end{equation}
with the convention, here and in the following, that
$\zrhh_{i i}^{N_i} \defd 1$ whenever $N_i = \zrhh_{i
  i}= 0$ (the reason is that the product originally is, to wit,
restricted to the terms with $N_i > 0$).

The discussion of the explicit form of the prior
$g(\zrh)\, \di\zrh$ is deferred to \sect~\ref{sec:prior}.
We shall first introduce on $\zsosett$ a suitable
coordinate system $(\zla_1, \dotsc, \zla_8) \equiv \zll
\in 
\RR^8$ so as to explicitly calculate the integrals. This
is done in the next section.

 
\section{Bloch vectors}
\label{sec:blochvectors}

In order to calculate the integrals required in the
state-assignment formula~\eqref{eq:rho_D_S_particular_exp}
we put a suitable coordinate system on $\zsosett$, so that
they ``translate'' as integrals in $\RR^8$. In
differential-geometrical terms, we choose a particular
chart on $\zsosett$ considered as a differentiable
manifold~\citep{kobayashietal1963,boothby1975_r1986,choquet-bruhatetal1977_r1996,marsdenetal1983_r2002,curtisetal1985,gallotetal1987,kennington2001_r2006}.

There exists an `Euler angle'
parametrisation~\citep{byrd1998,byrdetal2001,tilmaetal2002,tilmaetal2002b}
which maps $\zsosett$ onto a rectangular region of $\RR^8$
(modulo identification of some points). With this
parametrisation the integration limits of our
integrals become advantageously independent,
but the integrands ($\pf(D \cond \zrh)$ in particular) acquire
too complex a form.

For the latter reason we choose, instead, the
parametrisation studied by Byrd, Slater and
Khaneja~\citep{byrdetal2001,byrdetal2003},
Kimura~\citep{kimura2003} (see
also~\citep{kimuraetal2004}), and B\"ol\"ukba\c{s}\i\ and
Dereli~\citep{boeluekbasietal2006}, amongst others. The
functions to be integrated take simple polynomials or
exponentials forms. The integration limits are no longer
independent, though --- in fact, they are given in an
implicit form and will be accounted for by multiplying the
integrands by a characteristic function.

We follow Kimura's study~\cite{kimura2003} here, departing
from it on some definitions. All statistical operators of
a $d$-level quantum system can be written in the following
form~\cite{kimura2003} (see
also~\cite{byrdetal2001,byrdetal2003,kimuraetal2004}):
\begin{multline}
\label{eq:rhoexp}
\zrh = \zrh(\zll) = \frac{1}{d} \bm{I}_d + \frac{1}{2}
\sum_{j=1}^{n} \zla_j \zlgm_j,
\quad
(\zla_1,
\dotsc, \zla_n) \equiv \zll \in \mathbb{B}_n \subset \RR^n.
\end{multline} 
where $n \equiv d^2-1$ is the dimension of $\zsoset$, and
$\mathbb{B}_n$ is a compact convex subset of $\RR^n$. The
operators $\{\zlgm_j\}$ satisfy (1)
$\zlgm_j=\zlgm_j^{\dagger}$, (2) $\tr\zlgm_j=0$, (3)
$\tr(\zlgm_i\zlgm_j)=2\delt_{ij}$. Together with the
identity operator $\bm{I}_d$ they are generators of
$\mathrm{SU}(d)$, and in respect of the Frobenius
(Hilbert-Schmidt) inner product $\zlgm_i \cdot \zlgm_j
\defd \tr(\zlgm_i
\zlgm_j)$~\citep{hall2003_r2004} they also constitute a
complete orthogonal basis for the vector space of
Hermitean operators on a $d$-dimensional Hilbert space. In
fact, \eqn~\eqref{eq:rhoexp} is simply the decomposition
of the Hermitean operator $\zrh$ in terms of such a basis.
The vector $\zll \equiv (\zla_j)$ of coefficients in
equation~\eqref{eq:rhoexp} is uniquely determined by
$\zrh$:
\begin{equation}
\zla_j = \zla_j(\zrh)=\tr(\zlgm_j\,\zrh).\label{eq:x_of_rho}
\end{equation}
The operators $\{\zlgm_j\}$, being Hermitean, can also be
regarded as observables and then the equation above says
that the $(\zla_i)$ are the corresponding expectation
values in the state $\zrh$:
$\zla_j=\expe{\zlgm_j}_{\zrh}$~\citep{peres1995}.

A systematic construction of generators of
$\mathrm{SU}(d)$ which generalises the Pauli spin
operators is known (see
\eg~\cite{hioeetal1981,kimura2003}). In particular, for
$d=2$ they are the usual Pauli spin operators, and for
$d=3$ they are the Gell-Mann matrices (see
\eg~\citep{macfarlaneetal1968}). In the eigenbasis
$\set{\zone, \ztwo, \zthr}$ of the von~Neumann measurement
$\set{\ze_\mu}$ introduced in the previous section these
matrices assume the particular form
\begin{subequations}\label{eq:lambdaop}
  \begin{equation}
    \begin{gathered}
      \zlgm_{1} = \begin{pmatrix}
        0 & 1 & 0 \\
        1 & 0 & 0 \\
        0 & 0 & 0
      \end{pmatrix}, \quad \zlgm_{2} = \begin{pmatrix}
        0 & -\I & 0 \\
        \I & 0  & 0 \\
        0 & 0 & 0
      \end{pmatrix}, \quad \zlgm_{3} = \begin{pmatrix}
        1 & 0  & 0 \\
        0 & 0 & 0 \\
        0 & 0 & -1
      \end{pmatrix},
      \\
      \zlgm_{4} = \begin{pmatrix}
        0 & 0 & 1 \\
        0 & 0 & 0 \\
        1 & 0 & 0
      \end{pmatrix}, \quad \zlgm_{5} = \begin{pmatrix}
        0 & 0 & -\I \\
        0 & 0 & 0 \\
        \I & 0 & 0
      \end{pmatrix}, \quad \zlgm_{6} = \begin{pmatrix}
        0 & 0 & 0 \\
        0 & 0 & 1 \\
        0 & 1 & 0
      \end{pmatrix},
      \\
      \zlgm_{7} = \begin{pmatrix}
        0 & 0 & 0 \\
        0 & 0 & -\I \\
        0 & \I & 0
      \end{pmatrix}, \quad \zlgm_{8} = \frac{1}{\sqrt{3}}
      \begin{pmatrix}
        1 & 0 & 0 \\
        0 & -2 & 0 \\
        0 & 0 & 1
      \end{pmatrix}.
    \end{gathered}
  \end{equation}
  We see that our von~Neumann measurement corresponds to
  the observable
  \begin{equation}
    \zlgm_3\equiv \zone +0 \ztwo - \zthr,\label{eq:lambda_3-explic}
  \end{equation}
\end{subequations}
the measurement outcomes being associated with the particular values $1$,
$0$, and $-1$. These eigenvalues, however, are of no importance to us (they
will be more relevant in the companion paper~\citep{maanssonetal2007}).

For a three-level system, and in the eigenbasis $\set{\zone,
  \ztwo, \zthr}$, the operator $\zrh$
in~\eqref{eq:rhoexp} can thus be written in matrix form
as:
\begin{multline}
\label{eq:rhomatrix}
\zrh = \zrh(\zll)={}
\\ 
\begin{pmatrix}
                      \frac{1}{3}+\frac{1}{2}(\zla_3+\frac{1}{\sqrt{3}}\zla_8)
                    & \frac{1}{2}(\zla_1-\I\zla_2)  
                    & \frac{1}{2}(\zla_4-\I\zla_5) \\
                      \frac{1}{2}(\zla_1+\I\zla_2)  
                    & \frac{1}{3}-\frac{1}{\sqrt{3}}\zla_8
                    & \frac{1}{2}(\zla_6-\I\zla_7) \\
                      \frac{1}{2}(\zla_4+\I\zla_5) 
                    & \frac{1}{2}(\zla_6+\I\zla_7) 
                    & \frac{1}{3}+\frac{1}{2}(-\zla_3+\frac{1}{\sqrt{3}}\zla_8)
              \end{pmatrix}.
\end{multline} 
This matrix is Hermitean and has unit trace, so the
remaining condition for it to be a statistical operator is
that it be positive semi-definite (non-negative
eigenvalues). This is equivalent to two
conditions~\cite{kimura2003} for the coefficients $\zll$:
with our definitions of the Gell-Mann matrices, 
the first is
\begin{subequations}\label{eq:bv}
  \begin{equation}
    \label{eq:bvI}
    \zll^2 \equiv
    \tsum_{k=1}^{8} \zla_i^2 \le \frac{4}{3},
  \end{equation}
  which limits $\zll$ to be inside or on a ball of radius
  $2/\sqrt{3}$; the second is
  \begin{multline}
    8 - 18\zll^2 + 27\zla_3 \bigl(\zla_1^2 + \zla_2^2 -
    \zla_6^2 - \zla_7^2\bigr) -
    6\sqrt{3}\zla_8^3 + {}\\
    9\sqrt{3}\zla_8 \bigl[2\bigl(\zla_3^2 + \zla_4^2 +
    \zla_5^2\bigr) - \bigl(\zla_1^2
    + \zla_2^2 + \zla_6^2 + \zla_7^2\bigr)\bigr] + {}\\
    54 (\zla_1\zla_4\zla_6 + \zla_2\zla_4\zla_7 +
    \zla_2\zla_5\zla_6 - \zla_1\zla_5\zla_7) \ge
    0\label{eq:bvII}.
  \end{multline}
\end{subequations}
The set of all real vectors $\zll$ satisfying
conditions~\eqref{eq:bv} is called the `Bloch-vector
set' $\zxset$ of the three-level system:
\begin{equation}
  \label{eq:def_B}
  \zxset \defd \set{\zll \in \RR^8 \st
    \text{\eqref{eq:bv} hold}}.
\end{equation}
Since there is a bijective correspondence between $\zxset$
and $\zsosett$, we can parametrise the set of all
statistical operators $\zsosett$ by the set of all Bloch
vectors.\footnote{On some later occasions the terms
  `statistical operators' and `Bloch vectors' might be used
  interchangeably; but it should be clear from the context
  which one is really meant.} 

Both $\zsosett$ and $\zxset$ are convex
sets~\citep{valentine1964,gruenbaum1967_r2003,rockafellar1970,alfsen1971,broendsted1983,webster1994,segal1947,mackey1963,mielnik1968,mielnik1974,bloore1976},
and the maps
\begin{align}
  \label{eq:map_S-B}
  \zsosett &\to \zxset 
\quad\text{by} \quad
\zrh \mapsto \zll(\zrh)
\\
\intertext{given by~\eqref{eq:x_of_rho}, and its inverse}
  \zxset &\to \zsosett
\quad\text{by} \quad
\zll \mapsto \zrh(\zll)\label{eq:map_B-S}
\end{align}
given by~\eqref{eq:rhoexp} or~\eqref{eq:rhomatrix} are
convex isomorphisms, \ie\ they preserve convex
combinations:
\begin{align}
  \label{eq:conv_S-B}
\zll(\zcoa \zrha +  \zcob \zrhb) =
\zcoa \zll(\zrha) +  \zcob \zll(\zrhb),
\\
\zrh(\zcoa \zlla +  \zcob \zllb) =
\zcoa \zrh(\zlla) +  \zcob \zrh(\zllb),\label{eq:conv_B-S}
\end{align}
with $\zcoa, \zcob \ge 0$, $\zcoa + \zcob =1$. This fact
will be relevant for the discussion of the prior
distributions.

\begin{figure}[tb!]
\includegraphics[width=\columnwidth]{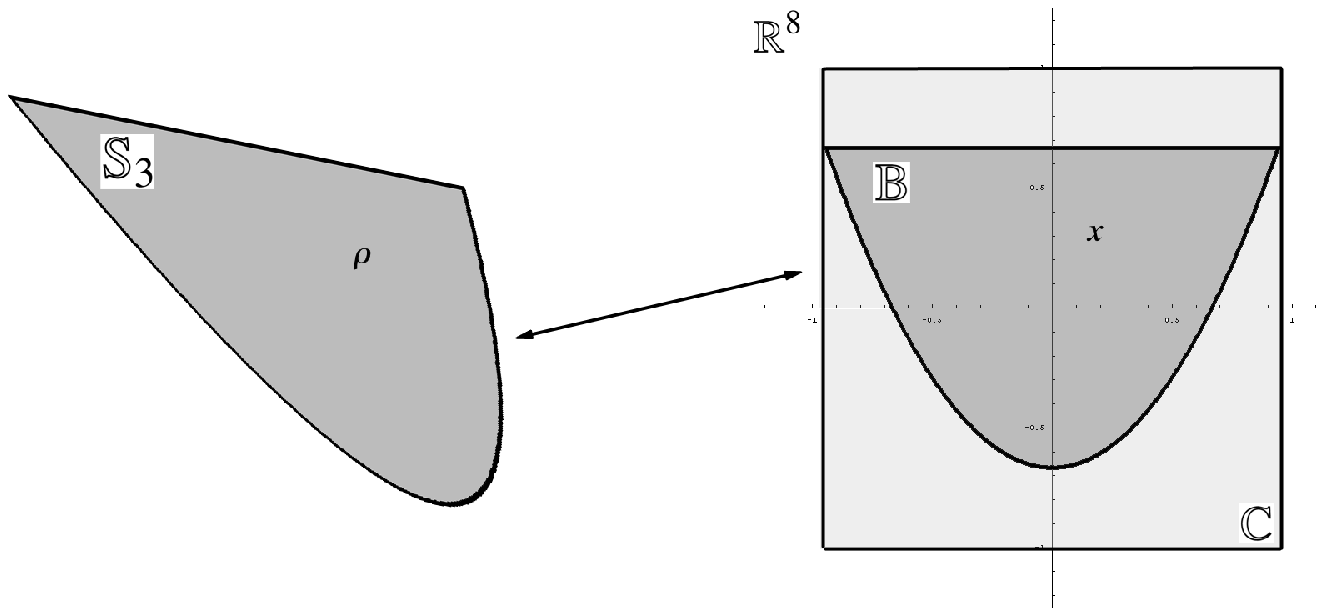}
\caption{Schematic illustration of the relations amongst $\zsosett$, $\zxset$, and $\zcset$.\label{fig:map}}
\end{figure}

It is useful to introduce the characteristic function
$\zll \mapsto \zchf(\zll)$ of the set $\zxset$:
\begin{equation}
  \label{eq:char_func_B}
  \zchf(\zll) \defd
  \begin{cases}
    1& \text{if $\zll \in \zxset$, \ie\ if~\eqref{eq:bv}
      hold},
\\
  0& \text{if $\zll \notin \zxset$, \ie\ if~\eqref{eq:bv}
      do not hold},
  \end{cases}
\end{equation}
and to consider the smallest eight-dimensional
rectangular region (or `orthotope'~\citep{gruenbaum1967_r2003})
$\zcset$ containing $\zxset$. As shown in the appendix,
$\zcset$ is
\begin{equation}
  \label{eq:def_C}
  \zcset \defd \clcl{-1,1}^7 \times
  \Bigl\lclose-\tfrac{2}{\sqrt{3}}, \tfrac{1}{\sqrt{3}}
  \Bigr\rclose
\supset \zxset.
\end{equation}
The relations amongst $\zsosett$, $\zxset$, and $\zcset$
are schematically illustrated in fig.~\ref{fig:map}. In
fig.~\ref{fig:sections} we can see some three-dimensional
sections (through the origin) of $\zxset$ --- and thus
of $\zsosett$ as well, in the sense of their isomorphism.

We are almost ready to write the integrals of
formula~\eqref{eq:rho_D_S_particular_exp} in coordinate
form, \ie\ as integrals over $\RR^8$. It only remains to
specify the volume element\footnote{An odd volume
  form~\citep{moser1965}\citep[\sect~IV.B.1]{choquet-bruhatetal1977_r1996}
  (see
  also~\citep{marsdenetal1983_r2002,kennington2001_r2006}).
  Recall that a metric structure is not required, only a
  differentiable one.} $\di\zrh$ in coordinate form. What
we shall do is in fact the opposite: we \emph{define}
$\di\zrh$ to be the volume element on $\zsosett$ which in
the coordinates $\zll$ is simply $\di\zll$. In
differential-geometrical terms, $\di\zrh$ is the
pull-back~\citep{choquet-bruhatetal1977_r1996,marsdenetal1983_r2002,curtisetal1985,gallotetal1987,kennington2001_r2006}
of $\di\zll$ induced by the map $\zrh \mapsto
\zll$:
\begin{equation}
  \label{eq:pull-back_drho}
  \di\zrh \mapsto \di\zll. 
\end{equation}

It is worth noting that this choice of volume element is
not arbitrary, but rather quite natural. On any
$n$-dimensional convex set $\zsconv$ we can define a
volume element which is canonical in respect of
$\zsconv$'s convex structure, as follows.
Consider any convex isomorphism $\zciso \colon \zsconv \to
\zrconv$ between $\zsconv$ and some subset $\zrconv
\subset \RR^n$. Consider the volume element on $\zrconv$
defined by
\begin{equation}
  \label{eq:can_vol_el}
  \zve \defd \di\zxx/\smallint_\zrconv \di\zxx,
\end{equation}
where $\di\zxx$ is the canonical volume element on
$\RR^n$. The pull-back $\zciso^*(\zve)$ of $\zve$ onto
$\zsconv$ then yields a volume element on the latter. It
is easy to see that the volume element thus induced (1) does
not depend on the particular isomorphism $\zciso$ (and set
$\zrconv$) chosen, 
since all such isomorphisms are related by affine
coordinate changes ($\zxx \mapsto \zA \zxx + \zb$, with
$\det\zA \ne 0$, $\zb \in \RR^n$); (2) is invariant in
respect of convex automorphisms of $\zsconv$; (3) assigns
unit volume to $\zsconv$, as clear from
\eqn~\eqref{eq:can_vol_el}. These properties make this
volume element canonical.\footnote{In measure-theoretic
  terms, we have the canonical measure $\zsubco \mapsto
  \zm[\zciso(\zsubco)]/\zm[\zciso(\zsconv)]$, where
  $\zsubco$ is a set of the appropriate $\sigma$-field of
  $\zsconv$ and $\zm$ is the Lebesgue measure on $\RR^n$.}

Since the parametrisation $\zsosett \to \zxset$ is a convex
isomorphism, we see that $\di\zrh$ as defined
in~\eqref{eq:pull-back_drho} is the canonical volume
element of $\zsosett$ in respect of its convex structure.

\medskip

We can finally write any integral over $\zsosett$ in
coordinate form. If $\zrh \mapsto f(\zrh)$ is an
integrable (possibly vector-valued) function over
$\zsosett$, its integral becomes
\begin{multline}
  \label{eq:integral_B_C}
\int_\zsosett f(\zrh)\, \di\zrh \equiv
\int_\zcset f[\zrh(\zll)]\, \zchf(\zll)\, \di\zll
\equiv{}\\
\int_{-1}^1 \di\zla_1\,
\dotsi
\int_{-1}^1\di\zla_7
\int_{-\frac{2}{\sqrt{3}}}^{\frac{1}{\sqrt{3}}}\di\zla_8
\, f[\zrh(\zll)]\, \zchf(\zll).
\end{multline}
This form is especially suited to numerical integration by
computer and we shall use it hereafter. We can thus
rewrite the state-assignment
formula~\eqref{eq:rho_D_S_particular_exp} for $\zrh_{D
  \land I}$ as:
\begin{equation}
  \label{eq:rho_D_particular_exp}
    \zrh_{D \land I}
    = \frac{ \int_\zcset \zrh(\zll)\, \Bigl[\tprod_{i=1}^3
      \zrhh_{i i}(\zll)^{N_i}\Bigr]\, \zchf(\zll)\, g(\zll)\, \di\zll 
}{
      \int_\zcset \Bigl[\tprod_{i=1}^3 \zrhh_{i i}(\zll)^{N_i}
\Bigr]\, \zchf(\zll)\, g(\zll)\, \di\zll }.
\end{equation}
Expanding the $\zrh(\zll)$ inside the integrals using
\eqn~\eqref{eq:rhoexp} (equivalent
to~\eqref{eq:rhomatrix}) we further obtain
\begin{equation}
\label{eq:genso}
\zrh_{D \land I} = \frac{1}{3} \bm{I}_3 + \frac{1}{2}
\sum_{j=1}^{8} \frac{L_j(\zNv, I)}{Z(\zNv, I)} \zlgm_j, 
\end{equation}
where
\begin{subequations}\label{eq:integralsz}
  \begin{align}
    \label{eq:integrals}
    L_j(\zNv, I) &\defd \int_\zcset 
    \zla_j\, \Bigl[\tprod_{i=1}^3 \rho_{ii}(\zll)^{N_i}\Bigr]\, 
    g(\zll)\,  \zchf(\zll)\,\di\zll,
\\
  \intertext{for $j=1, \dotsc, 8$, and}
    \label{eq:integralz}
    Z(\zNv, I) &\defd \int_\zcset 
    \Bigl[\tprod_{i=1}^3 \rho_{ii}(\zll)^{N_i}\Bigr]\, 
    g(\zll)\,  \zchf(\zll)\,\di\zll.
  \end{align}
\end{subequations}
We shall omit the argument `$(\zNv, I)$' from both $L_j$
and $Z$ when it should be clear from the context.

It is now time to discuss the prior plausibility
distributions adopted in our study.

\section{Prior knowledge}
\label{sec:prior}


The prior knowledge $I$ about the preparation is expressed
as a prior plausibility distribution $\pf(\zrh \cond I)\,
\di\zrh = g(\zrh)\, \di\zrh$. The last expression can be
interpreted, in measure-theoretic
terms~\citep{rudin1953_r1976,rudin1970}\citep[\sect~I.D]{choquet-bruhatetal1977_r1996}~\citep{fremlin2000_r2004}
(\cf\ also~\citep{kolmogorov1933_t1956,doob1996}), as
`$\mu(\di\zrh)$', where $\mu$ is a normalised measure; or
it can be simply interpreted, as we do here, as the
product of a generalised
function\footnote{\label{fn:egorov}We always use the term
  `generalised function' in the sense of
  Egorov~\citep{egorov1990}, whose theory is most general
  and nearest to the physicists' ideas and practice. \Cf\
  also Lighthill~\citep{lighthill1958_r1964},
  Colombeau~\citep{colombeau1984,colombeau1985,colombeau1992},
  and
  Oberguggenberger~\citep{oberguggenberger1992,oberguggenberger2001}.}
$g$ and the volume element
$\di\zrh$.\footnote{\label{fn:soffer}It is always
  preferable to write not only the plausibility density,
  but the volume element as well. The combined expression
  is thus invariant under parameter changes; this also
  helps not to fall into some pitfalls such as those
  discussed by Soffer and Lynch~\citep{sofferetal1999}.}
The two points of view are not mutually exclusive of
course, and these technical matters are only relatively
important since $\zsosett$ and the distributions we
consider are quite well-behaved objects (and the simple
Riemann integral suffices for our purposes).

We shall specify the plausibility distributions on
$\zsosett$ giving them directly in coordinate
form on $\zxset$ (with an abuse of notation for $g$):
\begin{equation}
  \label{eq:coord_g_gen}
\pf(\zll \cond I)\, \di\zll = 
g(\zll)\, \di\zll \defd  g[\zrh(\zll)]\, \di\zll.
\end{equation}

\begin{figure}[tb!]
\includegraphics[width=\columnwidth]{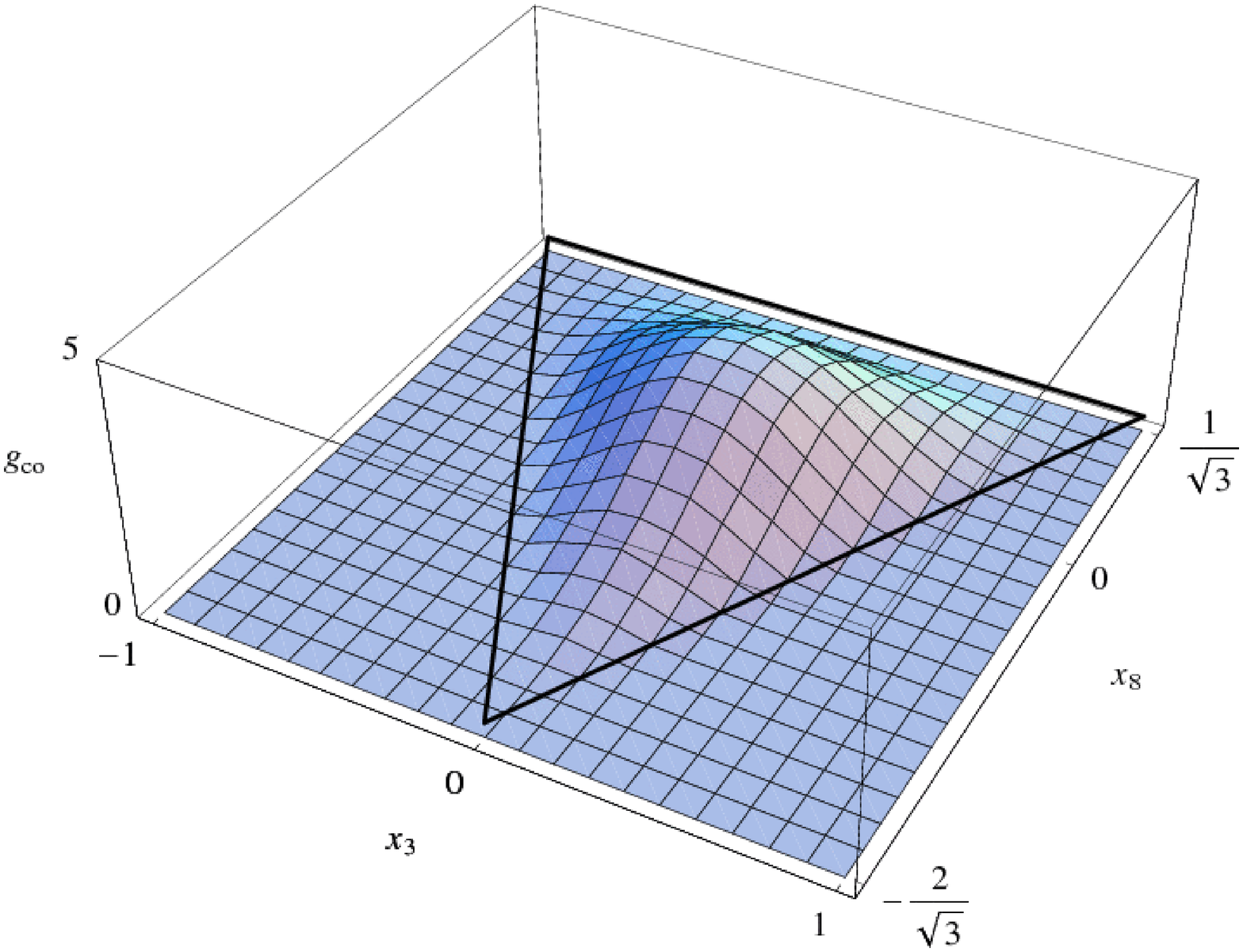}
\caption{\label{fig:priorc38}Graph of the constant prior's
  marginal density $(\zla_3,\zla_8) \mapsto \smallint
  \zgi\,
  \di\zla_1\,\di\zla_2\,\di\zla_4\,\di\zla_5\,\di\zla_6\,\di\zla_7$.
  The triangle represents the boundary of $\zxset$ in the
  $\origo\zla_3\zla_8$ plane (see
  \sect~\ref{sec:remaining}, and \cf\
  figs.~\ref{fig:Figure1}--\ref{fig:Figure5}).}
\end{figure}

The first kind of prior knowledge considered in our study,
$\zI$, has a constant density:
\begin{equation}
  \label{eq:first_prior_coo}
  \pf(\zll \cond \zI)\, \di\zll 
= \zgi(\zll)\, \di\zll
\propto \di\zll,
\end{equation}
the proportionality constant being given by the inverse of
the volume of $\zxset$. This distribution hence
corresponds to the canonical volume element (or the
canonical measure) discussed in the previous section. Thus
$\zI$ expresses somehow ``vague'' prior knowledge
(although we do not necessarily maintain that it be
``uninformative''). Fig.~\ref{fig:priorc38} shows the
marginal density of the coordinates $\zla_3$ and $\zla_8$
for this prior.
The state-assignment formula which makes use of this prior
assumes the simplified form
\begin{equation}
  \label{eq:rho_prior1}
    \zrh_{D \land \zI}
    = \frac{ \int_\zcset \zrh(\zll)\, \Bigl[\tprod_{i=1}^3
      \zrhh_{i i}(\zll)^{N_i}\Bigr]\, \zchf(\zll)\, \di\zll 
}{
      \int_\zcset \Bigl[\tprod_{i=1}^3 \zrhh_{i i}(\zll)^{N_i}
\Bigr]\, \zchf(\zll)\, \di\zll }.
\end{equation}

\begin{figure}[tb!]
\includegraphics[width=\columnwidth]{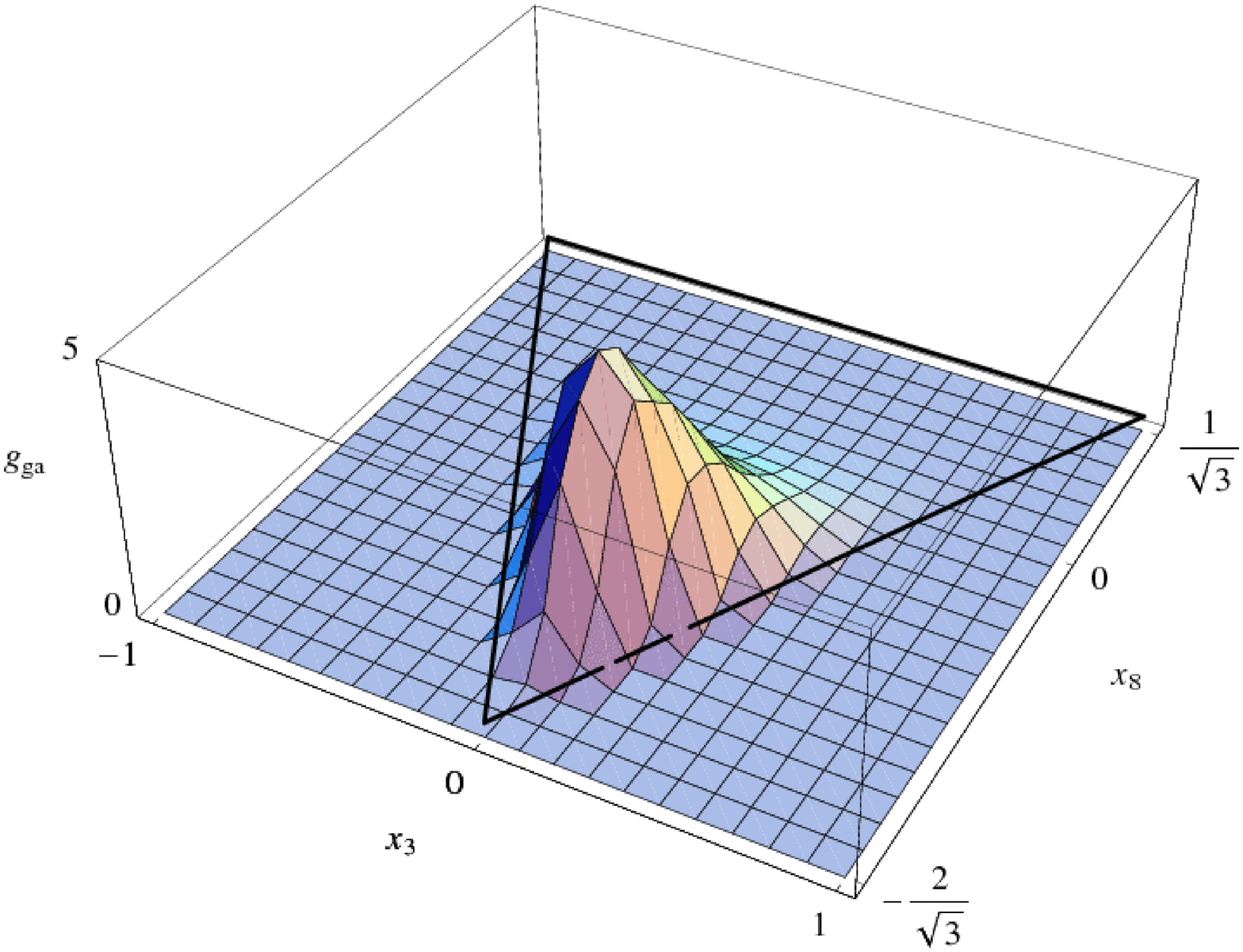}
\caption{\label{fig:priorg38}Graph of the Gaussian-like
  prior's marginal density $(\zla_3,\zla_8) \mapsto
  \smallint \zgj\,
  \di\zla_1\,\di\zla_2\,\di\zla_4\,\di\zla_5\,\di\zla_6\,\di\zla_7$.
  The triangle represents the boundary of $\zxset$ in the
  $\origo\zla_3\zla_8$ plane (see
  \sect~\ref{sec:remaining}, and \cf\
  figs.~\ref{fig:Figure1}--\ref{fig:Figure5}).}
\end{figure}

The second prior to be considered expresses somehow better
knowledge $\zJ$ of the possible preparation. In coordinate
form it is represented by the spherically symmetric
Gaussian-like distribution
\begin{multline}\label{eq:second_prior_coo}
    \pf(\zll \cond \zJ)\, \di\zll
= \zgj(\zll)\, \di\zll
\propto{}\\
    \exp\Biggl(
-\frac{\tr\bigl\{[\zrh(\zll)-\zrh(\zllc)]^2\bigr\}}{s^2}
\Biggr)\, \di\zll
    \equiv
\exp\Biggl[
\frac{(\zll - \zllc)^2}{2 s^2}
\Biggr]\, \di\zll,
\end{multline}
with
\begin{equation}
  \label{eq:detail_centre}
  \begin{aligned}
&    \zllc \defd (0, 0, 0, 0, 0, 0, 0, -2/\sqrt{3}),
    \quad\text{\ie,}\quad \zrh(\zllc) \equiv \ztwo, 
\\
&s = \frac{1}{2 \sqrt{2}}.
  \end{aligned}
\end{equation}
Regions in proximity of $\ztwo$ have greater plausibility,
and the plausibility of other regions decreases as their
``distance'' $\{\tr[\zrh(\zll)-\zrh(\zllc)]^2\}^{1/2}
\corr \abs{\zll - \zllc}$ from $\ztwo$ increases. The
parameter $s$ may be called the `breadth' of the
Gaussian-like function.\footnote{"Standard deviation"
  would be an improper name, \eg, since $s$ has not all
  the usual properties of a standard deviation. \Eg,
  although the Hessian determinant of the Gaussian-like
  density vanishes for $\abs{\zll - \zllc} =s$, the total
  plausibility within a distance $s$ from $\zllc$ is
  $0.0047$, not $0.00175$ as would be expected of an
  octavariate Gaussian distribution on
  $\RR^8$~\citep{chew1966}. This is simply due to the
  bounded ranges of the coordinates.} The marginal density
of the coordinates $\zla_3$ and $\zla_8$ for this prior is
shown in fig.~\ref{fig:priorg38}.
The state-assignment formula with the prior
knowledge $\zJ$ assumes the form
\begin{equation}
  \label{eq:rho_prior2}
    \zrh_{D \land \zJ}
    = \frac{ \int_\zcset \zrh(\zll)\, \Bigl[\tprod_{i=1}^3
      \zrhh_{i i}(\zll)^{N_i}\Bigr]\, 
\exp\Bigl[
\frac{(\zll - \zllc)^2}{2 s^2}
\Bigr]\,
\zchf(\zll)\, \di\zll 
}{
      \int_\zcset \Bigl[\tprod_{i=1}^3 \zrhh_{i i}(\zll)^{N_i}
\Bigr]\, 
\exp\Bigl[
\frac{(\zll - \zllc)^2}{2 s^2}
\Bigr]\,
\zchf(\zll)\, \di\zll }.
\end{equation}

In the following the function $g(\zll)$ will generically
stand for $\zgi(\zll)$ or $\zgj(\zll)$.

\section{Explicit calculation of the assigned statistical operator\label{sec:computation}}

We shall now calculate the \so\ given by~\eqref{eq:genso},
which means calculating the $L_j$ and $Z$ as given
in~\eqref{eq:integrals} and~\eqref{eq:integralsz}, for the
triples of absolute frequencies
\begin{align*}
  N&=1\text{:}&
&(1,0,0) \text{ and permutations thereof;}
\\
  N&=2\text{:}&
&(2,0,0), 
(1,1,0), 
\text{ and permutations;}
\\
N&=3\text{:}&
&(3,0,0), 
(2,1,0), 
(1,1,1),
\text{ and permutations;}
\\
N&=4,5,6,7\text{:}&
&(0,N,0),
\\
\intertext{with the prior distribution $\zgi(\zll)\, \di\zll$; and
the triples}
  N&=1\text{:}&
&(1,0,0), (0,1,0), (0,0,1);
\end{align*}
with the Gaussian-like prior distribution $\zgj(\zll)\, \di\zll$.

A combination of symmetries of $\zxset$ and
numerical integration is used to compute $L_j$ and $Z$.

\subsection{Deduction of some Bloch-vector parameters for
  some data via symmetry arguments}
\label{sec:symm}

The coefficients $L_j$ for $j = 1, 2, 4, 5, 6, 7$ can be
shown to vanish by symmetry arguments. Let us show
that $L_{1}=0$ in particular. Consider 
\begin{equation}
\label{eq:integral_1}
L_1 \equiv \int_\zcset 
\zla_1\, \Bigl[\tprod_{i=1}^3 \rho_{ii}(\zll)^{N_i}\Bigr]\, 
g(\zll)\, \zchf(\zll)\, \di\zll.
\end{equation}
The transformation
\begin{multline}
  \label{eq:transf_0}
 \zll \equiv (\zla_1,\zla_2,\zla_3,\zla_4,
\zla_5,\zla_6,\zla_7,\zla_8)
\mapsto{}
\\
\zllt \equiv (-\zla_1,\zla_2,\zla_3,\zla_4,
\zla_5,-\zla_6,\zla_7,-\zla_8)
\end{multline}
maps the domain $\zcset$ bijectively onto itself, and the
absolute value of its Jacobian determinant is equal to
unity. Under this transformation we have that
 \begin{subequations}\label{eq:trans_funct_1}
   \begin{align}
\zlat_1 &= -\zla_1,
\\
\zrh_{ii}(\zllt) &= \zrh_{ii}(\zll) \quad i = 1, 2, 3,
\\
g(\zllt) &= g(\zll) \quad\text{(for both $g=\zgi, \zgj$)},
\\
\zchf(\zllt) &= \zchf(\zll).
   \end{align}
 \end{subequations}
Applying the formula for the change of
variables~\citep{schwartz1954,lax1999} to
\eqref{eq:integral_1}, using the symmetries above, and
renaming dummy integration variables we obtain
\begin{align}
\label{eq:transf_integrals}
\begin{split}
  L_1 &= \int_\zcset \zla_1\, \Bigl[\tprod_{i=1}^3
  \rho_{ii}(\zll)^{N_i}\Bigr]\, g(\zll)\, \zchf(\zll)\,
  \di\zll, 
\\
&= -\int_\zcset\zla_1\, \Bigl[\tprod_{i=1}^3
  \rho_{ii}(\zll)^{N_i}\Bigr]\, g(\zll)\, \zchf(\zll)\,
  \di\zll,
\end{split}
\\
\therefore  L_1 &=0.
\end{align}

Similarly one can show that
$L_{2}$, $L_{4}$, $L_{5}$, $L_{6}$, $L_{7}$ are all zero by 
changing the signs of the triplets
$(\zla_2,\zla_5,\zla_7)$, $(\zla_1,\zla_5,\zla_6)$,
$(\zla_2,\zla_4,\zla_6)$, $(\zla_2,\zla_4,\zla_6)$,
$(\zla_2,\zla_5,\zla_7)$,
respectively.


The assigned \so\ hence corresponds to the Bloch vector
$(0,0,L_3/Z,0,0,0,0,L_8/Z)$, for all triples of absolute
frequencies $\zNv$ and both kinds of prior knowledge. \Ie\ it has, in the
eigenbasis $\set{\zone, \ztwo, \zthr}$, the diagonal
matrix form
\begin{equation}
\label{eq:rhomatrix_diag}
\zrh_{D \land I}=
\begin{pmatrix}
                      \frac{1}{3}+\frac{L_3+ L_8/\sqrt{3}}{2 Z}
                    & 0
                    &0 \\
                     0  
                    & \frac{1}{3}-\frac{L_8}{\sqrt{3} Z}
                    & 0 \\
                     0
                    & 0
                   & \frac{1}{3}+\frac{-L_3+
                      L_8/\sqrt{3}}{2 Z}
               \end{pmatrix}
\end{equation} 
(note that $L_{3,8}$ and $Z$ still depend on $\zNv$ and
$I$).


Two further changes of variables --- both with unit
Jacobian determinant and mapping $\zxset$ 1-1 onto itself
--- can be used to reduce the calculations for some
absolute-frequency triples $(N_1, N_2, N_3)$ to the
calculation of other ones, with a reasoning similar to
that of the preceding section.

The first is
\begin{multline}
  \label{eq:transf_1}
\zll \equiv (\zla_1,\zla_2,\zla_3,\zla_4,
\zla_5,\zla_6,\zla_7,\zla_8)
\mapsto{}
\\
\zllt \equiv 
(\zla_6, \zla_7,-\zla_3,\zla_4,-\zla_5,\zla_1,\zla_2,
\zla_8),
\end{multline}
under which, in particular,
\begin{equation}
  \label{eq:sym1_rho}
  \zrh_{11}(\zllt)=
  \zrh_{33}(\zll),
\quad
  \zrh_{33}(\zllt)=
  \zrh_{11}(\zll),
\quad
  \zrh_{22}(\zllt)=
  \zrh_{22}(\zll).
\end{equation}
From \eqns~\eqref{eq:integralsz} it follows that
\begin{subequations}\label{eq:sym1_LZ}
  \begin{align}
  L_3(N_3, N_2, N_1) &= -L_3(N_1, N_2, N_3),
\\
  L_8(N_3, N_2, N_1) &= L_8(N_1, N_2, N_3),
\\
  Z(N_3, N_2, N_1) &= Z(N_1, N_2, N_3),
  \end{align}
\end{subequations}
for both prior distributions $\zgi$ and $\zgj$.

The second change of variables is an anti-clockwise
rotation of the plane $(\zla_3,\zla_8)$ by an angle
$2\pi/3$ accompanied by permutations of the other
coordinates:
\begin{multline}
  \label{eq:transf_2}
 (\zla_1,\zla_2,\zla_3,\zla_4,
\zla_5,\zla_6,\zla_7,\zla_8)
\mapsto{}\\
\biggl(\zla_7,\zla_6,-\frac{\zla_3 +
  \sqrt{3}\zla_8}{2},\zla_2,
\zla_1,\zla_4,\zla_5,\frac{\sqrt{3}\zla_3
  -\zla_8}{2}\Biggr),
\end{multline}
under which, in particular,
\begin{equation}
  \label{eq:sym2_rho}
  \zrh_{11}(\zllt)=
  \zrh_{22}(\zll),
\quad
  \zrh_{22}(\zllt)=
  \zrh_{33}(\zll),
\quad
  \zrh_{33}(\zllt)=
  \zrh_{11}(\zll),
\end{equation}
leading to
\begin{subequations}\label{eq:sym2_LZ}
\begin{align}
  \begin{split}
    L_3[(N_2, N_3, N_1), \zI] &= -\frac{1}{2} L_3[(N_1,
    N_2, N_3), \zI] -{}
\\
&\qquad\qquad\frac{\sqrt{3}}{2} L_8[(N_1, N_2,
    N_3), \zI],
  \end{split}
\\
\begin{split}
  L_8[(N_2, N_3, N_1), \zI] &= \frac{\sqrt{3}}{2}L_3[(N_1, N_2,
  N_3), \zI] -{}
\\
&\qquad\qquad\frac{1}{2}L_8[(N_1, N_2, N_3), \zI],
\end{split}
\\
  Z[(N_2, N_3, N_1), \zI] &= Z[(N_1, N_2, N_3), \zI].
  \end{align}
\end{subequations}
Note that the formulae from this transformation holds
only for the constant prior $\zgi$. 

\medskip

From~\eqref{eq:sym1_LZ} we see that, for
\emph{both} priors, $L_3$ vanishes for all triples of the
form $(n, N-2n,n)$ for some positive integer $n\le N/2$,
in particular for $(0,N,0)$ and $(n, n, n)$. In the last
case $L_8=0$ as well --- though only for the constant
prior $\zgi$ ---, as can be deduced
from~\eqref{eq:sym1_LZ} and~\eqref{eq:sym2_LZ}.

In the case of the prior knowledge $\zI$, it is easy to
realise that, repeatedly applying the two transformations
above, one can derive the values of $L_3$, $L_8$, and $Z$
for all triples $(N_1, N_2, N_2)$ from the values for the
triples with $N_2 \ge N_1 \ge N_3$ only.

\subsection{Numerical calculation 
for the remaining cases}
\label{sec:remaining}

No other symmetry arguments seem available to derive
$L_3$, $L_8$, and $Z$ for the remaining cases. In fact
$L_3$, $L_8$ are in general non-zero ($Z$ can never
vanish, its integrand being positive and never identically
naught). It is very difficult --- impossible perhaps?\ ---
to calculate the corresponding integrals analytically
because of the complicated shape of $\zxset$. Therefore we
have resorted to numerical integration, using the
quasi-Monte~Carlo integration algorithms provided by
Mathematica~5.2.\footnote{The programmes are available
  upon request.}

The resulting Bloch vectors for the constant prior $\zgi
\, \di\zll$ are shown for $N=1,2,3$ in
figs.~\ref{fig:Figure1}, \ref{fig:Figure2},
and~\ref{fig:Figure3} respectively. We have included in
fig.~\ref{fig:Figure1} the case $N=0$ --- \ie, no data ---
corresponding to the \so\ $\zrh_{\zI}$ that encodes the
prior knowledge $\zI$. In fig.~\ref{fig:Figure4} we have
plotted the Bloch vectors corresponding to triples of the
form $(N_1,N_2,N_3)=(0,0,N)$ for $N=1,\dotsc,7$.

The cases $N=0$ and $N=1$ for the Gaussian-like prior
$\zgj\, \di\zll$ are shown in fig.~\ref{fig:Figure5}. The
case $N=0$ corresponds to the \so\ $\zrh_{\zJ}$ encoding
the prior knowledge $\zJ$.

The large triangle in the figures is the two-dimensional
section of the set $\zxset$ along the plane
$\origo\zla_3\zla_8$. It can, of course, also be
considered as a section of the set of \so s $\zsosett$.
This section contains the eigenprojectors $\zone$,
$\ztwo$, $\zthr$, which are the vertices of the triangle,
as indicated. The assigned \so s, for all data and priors
considered in this study, also lie on this triangle since
they are mixtures of the eigenprojectors, as we found in
\sect~\ref{sec:symm}, \eqn~\eqref{eq:rhomatrix_diag}. They
are represented by points labelled with the respective
data triples. The points have planar coordinates
$\bigl(L_3(\zNv, I)/Z(\zNv, I), L_8(\zNv, I)/Z(\zNv,
I)\bigr)$.

The numerical-integration uncertainties $\zux$ and $\zuy$,
for $L_3/Z$ and $L_8/Z$ respectively, specified in the
figures' legends, vary from $\pm 0.0025$ for the triplets
with $N=2$ to $\pm 0.015$ for various other triplets.
Numerical integration has also been performed for those
quantities that can be determined analytically
(\sect~\ref{sec:symm}) --- like $L_3(0,N,0)/Z(0,N,0)$ \eg\
---, and the numerical results agree, within the
uncertainties, with the analytical ones.

A trade-off between, on the one hand, calculation time
and, on the other, accuracy of the result was necessary.
The accuracy parameters to be inputted onto the
integration routine were determined by previous rough
numerical estimations of the results; in some cases an
iterative process of this kind was adopted. The
calculation of the \so\ for a given triple of absolute
frequencies $\zNv$ took from three to one hundred minutes,
depending on the accuracy required and the complexity of
the integrands.

\begingroup
\begin{table*}[p!]
\begin{tabular}{c@{\hspace{2em}}c}
  \begin{tabular}[t]{l}
  \begin{tabular}{|l|}
\hline
\fbox{$\zI$, $N=0$ (no data):} 
\\
$\zrh_{(000), \zI} =   \begin{pmatrix}
                          1/3 & 0 & 0 \\
                          0 & 1/3 & 0 \\
                          0 & 0 & 1/3  
                        \end{pmatrix} $
\\ NB: This \so\ encodes the prior knowledge $\zI$
\\ \hline
  \end{tabular}
\\ \\ \\
  \begin{tabular}{|l|}
\hline
\fbox{$\zI$, $N=1$:} \\
$\zrh_{(010), \zI} =   \begin{pmatrix}
                          0.300 \pm 0.001\footnotemark[1] & 0 & 0 \\
                          0 & 0.399 \pm 0.003\footnotemark[1] & 0 \\
                          0 & 0 & 0.300 \pm 0.001\footnotemark[1]
                        \end{pmatrix} $
\\ cases $(100)$ and $(001)$ obtained by permutation
\\ \hline
  \end{tabular}
\\ \\ \\
  \begin{tabular}{|l|}
\hline
\fbox{$\zI$, $N=2$:} \\
 $\zrh_{(020), \zI} =   \begin{pmatrix}
                          0.2735 \pm 0.0007\footnotemark[1] & 0 & 0 \\
                          0 & 0.453 \pm 0.001\footnotemark[1] & 0 \\
                          0 & 0 &  0.2735 \pm 0.0007\footnotemark[1]
                        \end{pmatrix}$
\\ $\zrh_{(101), \zI} =   \begin{pmatrix}
                          0.3642 \pm 0.0007\footnotemark[1] & 0 & 0 \\
                          0 & 0.272 \pm 0.001\footnotemark[1] & 0 \\
                          0 & 0 & 0.3642  \pm 0.0007\footnotemark[1]
                        \end{pmatrix}$
\\ other cases obtained by permutation
\\ \hline
  \end{tabular}
\\ \\ \\
   \begin{tabular}{|l|}
\hline
\fbox{$\zI$, $N=3$:} \\
 $\zrh_{(030), \zI} =   \begin{pmatrix}
                          0.249 \pm 0.001\footnotemark[1] & 0 & 0 \\
                          0 & 0.502 \pm 0.003\footnotemark[1] & 0 \\
                          0 & 0 & 0.249 \pm 0.001\footnotemark[1] 
                        \end{pmatrix}$
\\ $\zrh_{(021), \zI} =\footnotemark[2]   \begin{pmatrix}
                          0.333 \pm 0.004\footnotemark[1] & 0 & 0 \\
                          0 & 0.418 \pm 0.003\footnotemark[1] & 0 \\
                          0 & 0 & 0.249 \pm 0.004\footnotemark[1]
                        \end{pmatrix}$
\\ $\zrh_{(111), \zI} =   \begin{pmatrix}
                          1/3 & 0 & 0 \\
                          0 & 1/3 & 0 \\
                          0 & 0 & 1/3 
                        \end{pmatrix}$
\\  other cases obtained by permutation
\\ \hline
\end{tabular}
\\ \\ \\
\begin{tabular}{|l|}
\hline
\fbox{$\zI$, $N=4$:} \\
$\zrh_{(040), \zI} =   \begin{pmatrix}
                          0.230 \pm 0.004\footnotemark[1] & 0 & 0 \\
                          0 & 0.541 \pm 0.009\footnotemark[1] & 0 \\
                          0 & 0 & 0.230 \pm 0.004\footnotemark[1]
                        \end{pmatrix}$
\\ cases $(400)$ and $(004)$ obtained by permutation
\\ \hline
  \end{tabular}
\end{tabular}
&
\begin{tabular}[t]{l}
\begin{tabular}{|l|}
\hline
\fbox{$\zI$, $N=5$:} \\
 $\zrh_{(050), \zI} =   \begin{pmatrix}
                          0.215 \pm 0.004\footnotemark[1] & 0 & 0 \\
                          0 & 0.571  \pm 0.009\footnotemark[1] & 0 \\
                          0 & 0 &  0.215  \pm 0.004\footnotemark[1]
                        \end{pmatrix}$
\\  cases $(500)$ and $(005)$ obtained by permutation
\\ \hline
\end{tabular}
\\ \\ \\
\begin{tabular}{|l|}
\hline
\fbox{$\zI$, $N=6$:} \\
 $\zrh_{(060), \zI} =   \begin{pmatrix}
                          0.201  \pm 0.004\footnotemark[1] & 0 & 0 \\
                          0 & 0.598 \pm 0.009\footnotemark[1] & 0 \\
                          0 & 0 & 0.201  \pm 0.004\footnotemark[1] 
                        \end{pmatrix}$
\\  cases $(600)$ and $(006)$ obtained by permutation
\\ \hline
\end{tabular}
\\ \\ \\
\begin{tabular}{|l|}
\hline
\fbox{$\zI$, $N=7$:} \\
 $\zrh_{(070), \zI} =   \begin{pmatrix}
                          0.191 \pm 0.004\footnotemark[1] & 0 & 0 \\
                          0 & 0.619 \pm 0.009\footnotemark[1] & 0 \\
                          0 & 0 & 0.191 \pm 0.004\footnotemark[1]
                        \end{pmatrix}$
\\  cases $(700)$ and $(007)$ obtained by permutation
\\ \hline
  \end{tabular}
\\ \\ \\
  \begin{tabular}{|l|}
\hline
\fbox{$\zJ$, $N=0$ (no data):} \\
$\zrh_{(000), \zJ} =   \begin{pmatrix}
                          0.195 \pm 0.004\footnotemark[1] & 0 & 0 \\
                          0 &  0.609 \pm 0.009\footnotemark[1] & 0 \\
                          0 & 0 &  0.195 \pm 0.004\footnotemark[1]
                        \end{pmatrix}$
\\ NB: This \so\ encodes the prior knowledge $\zJ$
\\ \hline
  \end{tabular}
\\ \\ \\
  \begin{tabular}{|l|}
\hline
\fbox{$\zJ$, $N=1$:} \\
 $\zrh_{(010), \zJ} =   \begin{pmatrix}
                          0.180 \pm 0.004\footnotemark[1]& 0 & 0 \\
                          0 & 0.640 \pm 0.009\footnotemark[1] & 0 \\
                          0 & 0 & 0.180 \pm 0.004\footnotemark[1]
                        \end{pmatrix}$
 \\ $\zrh_{(001), \zJ} =   \begin{pmatrix}
                          0.239 \pm 0.006\footnotemark[1] & 0 & 0 \\
                          0 & 0.575 \pm 0.009\footnotemark[1] & 0 \\
                          0 & 0 & 0.186 \pm 0.006\footnotemark[1] 
                        \end{pmatrix} $
 \\ $\zrh_{(100), \zJ} =   \begin{pmatrix}
                          0.186 \pm 0.006\footnotemark[1] & 0 & 0 \\
                          0 & 0.575 \pm 0.009\footnotemark[1] & 0 \\
                          0 & 0 &  0.239 \pm 0.006\footnotemark[1]
                        \end{pmatrix} $
\\ \hline 
\end{tabular}
\\
  \parbox{0.9\columnwidth}{
\begin{flushleft}
\footnotesize \footnotemark[1]Note that only two of the
    three uncertainties of the diagonal elements are
    independent; see \sect~\ref{sec:remaining}.
\\
\footnotesize \footnotemark[2]This has been computed from
the average of the cases $(021)$ and $(120)$
(appropriately permuted).
\end{flushleft}
}
\end{tabular}
\end{tabular}
  \caption{Statistical operators assigned for the various
    absolute-frequency data and priors considered in this
    study. \Cf\ figs.~\ref{fig:Figure1}--\ref{fig:Figure5}.}
  \label{tab:SO}
\end{table*}
\endgroup

The \so s encoding the various kinds of data and prior
knowledge are given in explicit form in
table~\ref{tab:SO}. Note that the uncertainties for the
\so s should be written as $\zux \zlgm_3/2 + \zuy
\zlgm_8/2$ (\cf\ \eqn~\eqref{eq:genso}); however, we
adopted a more compact notation in the table (see footnote
\textit{a} there).

The results for $N=2$ and $N=3$ show an intriguing
feature, immediately apparent in figs.~\ref{fig:Figure2}
and~\ref{fig:Figure3}: the computed Bloch vectors seem to
maintain the convex structure of the respective data. What
we mean is the following. For given $N$, the set  of possible
triples of absolute frequencies $(N_1, N_2, N_3)$ has a
natural convex structure with the extreme points
$(N,0,0)$, $(0,N,0)$, and $(0,0,N)$:
\begin{multline}
  \label{eq:conv_N123}
  (N_1, N_2, N_3) \equiv (f_1 N, f_2 N, f_3 N) ={}\\
f_1 (N,0,0) + 
f_2 (0,N,0) + 
f_3 (0,0,N),
\end{multline}
where we have introduced the relative frequencies $f_i
\defd N_i/N$. Denote the Bloch vector corresponding to the
triple $(N_1, N_2, N_3)\equiv (f_1 N, f_2 N, f_3 N)$ by
\begin{multline}
  \label{eq:bloch_v_short}
\zP(N_1, N_2, N_3) \defd{}\\
\bigl(0,0,L_3(\zNv,
\zI)/Z(\zNv, \zI), 0,0,0,0,L_8(\zNv, \zI)/Z(\zNv, \zI)\bigr).
\end{multline}
These Bloch vectors (and hence the \so s)
seem, from figs.~\ref{fig:Figure2}
and~\ref{fig:Figure3}, to respect the same convex
combinations as their respective triples:
\begin{equation}
  \label{eq:bloch_conv}
   \zP (f_1 N, f_2 N, f_3 N) \approx 
f_1 \zP(N,0,0) + 
f_2 \zP(0,N,0) + 
f_3 \zP(0,0,N).
\end{equation}
In terms of the integrals~\eqref{eq:integralsz} defining
$L_3$, $L_8$, $Z$, and using~\eqref{eq:rhomatrix}
or~\eqref{eq:rhomatrix_diag}, the seeming equation above becomes
\begin{widetext}
  \begin{multline}
    \label{eq:int_conv}
    \frac{ \int_\zxset 
      \zla_j\, \Bigl(\frac{1}{3} + \frac{\zla_3}{2} +
      \frac{\zla_8}{2 \sqrt{3}}\Bigr)^{f_1 N}\,
      \Bigl(\frac{1}{3} -
      \frac{\zla_8}{\sqrt{3}}\Bigr)^{f_2 N}\,
      \Bigl(\frac{1}{3} - \frac{\zla_3}{2} +
      \frac{\zla_8}{2 \sqrt{3}}\Bigr)^{f_3 N}\, 
      \di\zll
    }{
      \int_\zxset 
      \Bigl(\frac{1}{3} + \frac{\zla_3}{2} +
      \frac{\zla_8}{2 \sqrt{3}}\Bigr)^{f_1 N}\,
      \Bigl(\frac{1}{3} -
      \frac{\zla_8}{\sqrt{3}}\Bigr)^{f_2 N}\,
      \Bigl(\frac{1}{3} - \frac{\zla_3}{2} +
      \frac{\zla_8}{2 \sqrt{3}}\Bigr)^{f_3 N}\, 
      \di\zll
    }
 \approx{} \\  
f_1\frac{
      \int_\zxset 
      \zla_j\, \Bigl(\frac{1}{3} + \frac{\zla_3}{2} +
      \frac{\zla_8}{2 \sqrt{3}}\Bigr)^N\,
      \di\zll
    }{
      \int_\zxset 
      \Bigl(\frac{1}{3} + \frac{\zla_3}{2} +
      \frac{\zla_8}{2 \sqrt{3}}\Bigr)^N\,
      \di\zll
    }
    + 
f_2\frac{
      \int_\zxset 
      \zla_j\, \Bigl(\frac{1}{3} -
      \frac{\zla_8}{\sqrt{3}}\Bigr)^N\,
      \di\zll
    }{
      \int_\zxset 
      \Bigl(\frac{1}{3} -
      \frac{\zla_8}{\sqrt{3}}\Bigr)^N\,
      \di\zll
    }
    + 
    f_3\frac{
      \int_\zxset 
      \zla_j\, \Bigl(\frac{1}{3} - \frac{\zla_3}{2} +
      \frac{\zla_8}{2 \sqrt{3}}\Bigr)^N\,
      \di\zll
    }{
      \int_\zxset  \Bigl(\frac{1}{3} - \frac{\zla_3}{2} +
      \frac{\zla_8}{2 \sqrt{3}}\Bigr)^N\,
      \di\zll
    },\quad j=3, 8,
  \end{multline}
\end{widetext}
a remarkable expression. Does it hold exactly? We have not tried to prove
or disprove its analytical validity, but it surely deserves further
investigation. [\emph{Post scriptum:} Slater, using cylindrical algebraic
decomposition~\citep{arnonetal1984,arnonetal1984b,jirstrand1995} and a
parametrisation by Bloore~\citep[\cf][]{slater2007}, has confirmed that
\eqn~\eqref{eq:int_conv} holds exactly. In fact, he has remarked that the
some of the integrals, here numerically calculated, can be solved
analytically by his approach.]

\section{Taking account of the uncertainties in the
  detection of outcomes}
\label{sec:measerr}

Uncertainties are normally to be found in one's
measurement data, and need to be taken into account in the
state-assignment procedure. For frequency data the
uncertainty can stem from a combination of ``over-counting'',
\ie\ the registration (because of background noise \eg)\ of
some events as outcomes when there are in fact none, and
``under-counting'', \ie\ the failure (because of  detector
limitations, \eg)\ to register some outcomes.

Let us model the measurement-data uncertainty as follows,
for definiteness. We say that the plausibility of
registering the ``event'' `$i$'  when the outcome `$\mu$' is
obtained is
\begin{equation}
  \label{eq:p_unc_m-i}
  \pr(\text{`$i$'} \cond \text{`$\mu$'} \land I) = h(i \cond \mu).
\end{equation}
The event `$i$' belongs to some given set that may include
such events as \eg\ the `null', no-detection event; the
number of events need not be the same as the number of
outcomes. The model formalised in the equation above
suffices in many cases. Other models could take into
account, \eg\ ``non-local'' or memory effects, so that the
plausibility of an event could depend on a set of previous
or simultaneous outcomes. We thus definitely enter the
realm of communication
theory~\citep{shannon1948,shannon1949,middleton1960,csiszaretal1981,coveretal1991}
(see also~\citep{helstrom1967,helstrom1976}).

Given the preparation represented by the \so\ $\zrh$, and
the \povm\ $\set{\zee_\mu}$ representing the measurement
with outcomes $\set{\text{`$\mu$'}}$, the plausibility of
registering the event `$i$' in a measurement instance is,
by the rules of plausibility theory,\footnote{It is
  assumed that knowledge of the state is redundant in the
  plausibility assignment of the event `$i$' when the
  outcome is already known.}
\begin{equation}
  \label{eq:plaus_ev_i}
  \pf(i \cond \zrh) =
\tsum_\mu  \pf(i \cond \mu)\, \pf(\mu \cond \zrh) =
\tsum_\mu  h(i \cond \mu)\, \tr(\zee_\mu \zrh).
\end{equation}
This marginalisation could be carried over to the
state-assignment formulae already discussed in
\sect~\ref{sec:scenario}, and the formulae thus obtained
would take into account the outcome-registration
uncertainties.

However, it is much simpler to introduce a new \povm\
$\set{\zen_i}$ defined by
\begin{equation}
  \label{eq:def_F}
  \zen_i \defd \tsum_\mu h(i \cond \mu) \zee_\mu,
\end{equation}
so that the plausibilities $\pf(i \cond \zrh)$ in
\eqn~\eqref{eq:plaus_ev_i} can be written, by the
linearity of the trace,
\begin{equation}
  \label{eq:plaus_ev_i_new}
  \pf(i \cond \zrh) = \tr(\zen_i \zrh).
\end{equation}
In the state assignment we can simply use the new
\povm, which includes the outcome-registration
uncertainties, in place of the old one. The last procedure
is also more in the spirit of quantum mechanics: it is
analogous to the use of the \so\ $p_1 \zrh_1 + p_2 \zrh_2$
when we are unsure (with plausibilities $p_1$ and $p_2$)
about whether $\zrh_1$ or $\zrh_2$ holds. \Ie, we can
``mix'' \povmm\ elements just like we mix \so s. In fact,
we could even mix, with a similar procedure, whole \povm s
--- a procedure which would represent the fact that there
are uncertainties in the identification not only of the
outcomes, but of the whole measurement procedure as well.
See Peres' partially related
discussion~\citep{peres2002}.

\section{Large-$N$ limit}\label{sec:N_infinity}

\subsection{General case}
\label{sec:N_large_gen_case}

Let us briefly consider the case of data with very large $N$. We summarise
some results obtained in~\citep{portamana2007}. Mathematically we want to
see what form the state-assignment formulae take in the limit $N \to
\infty$. Consider a sequence of data sets $\set{D_N}_{N=1}^\infty$. Each
$D_N$ consists in some knowledge about the outcomes of $N$ instances of the
same measurement. The latter is represented by the \povm\ $\set{\ze_i}$.
The plausibility distribution for the outcomes, given the preparation
$\zrh$, is
\begin{equation}
  \label{eq:plaus_out_N}
  \zqq(\zrh)
  \defd \bigl(\zq_i(\zrh)\bigr)\quad \text{with} \quad \zq_i(\zrh) = \tr(\ze_i \zrh).
\end{equation}

Let us consider more precisely the general situation in
which each data set $D_N$ consists in the knowledge that
the relative frequencies $\zf \equiv(f_i) \defd (N_i/N)$
lie in a region $\zzf_N$ (with non-empty interior and whose
boundary has measure zero in respect of the prior
plausibility measure). Such kind of data
arise when the registration of measurement outcomes is
affected by uncertainties and is moreover
``coarse-grained'' for practical purposes, so that not
precise frequencies are obtained but rather a region ---
like $\zzf_N$ --- of possible ones.

For each data set we then have a resulting posterior
distribution for the \so s,
\begin{multline}
  \label{eq:post_N}
  \pf(\zrh \cond  D_N \land I)\, \di\zrh
=  \pf[\zrh \cond ( \zf \in \zzf_N) \land I]\, \di\zrh
={}\\
 \frac{ \pf(\zf \in \zzf_N
      \cond \zrh)\, \pf(\zrh \cond I)\, \di\zrh } {\int_{\zsoset}
      \pf(\zf \in \zzf_N \cond \zrh)\, \pf(\zrh \cond I)\, \di\zrh}.
\end{multline}
and an associated \so\ $\zrh_{D_N \land I} \defd \smallint
\zrh \,\pf(\zrh \cond D_N \land I)\, \di\zrh$. 

Assume that the sequence $\set{\zzf_N}_{N=1}^\infty$ of such frequency
regions converges (in a topological sense specified
in~\citep{portamana2007}) to a region $\zzfi$ (also with non-empty interior
and with boundary of measure zero). We shall see later what happens when
such a region shrinks to a single point, \ie\ when the uncertainties
becomes smaller and smaller. In~\citep{portamana2007} it is shown, using
some theorems in Csisz\'ar~\citep{csiszar1984} and Csisz\'ar and
Shields~\citep{csiszaretal2004b}, that
\begin{multline}
  \label{eq:post_N}
  \pf(\zrh \cond  D_N \land I)\, \di\zrh
\propto 
\begin{cases}
  0,& \text{if $\zqq(\zrh) \not\in \zzfi$},
\\
 \pf(\zrh \cond I)\, \di\zrh,
& \text{if  $\zqq(\zrh) \in \zzfi$},
\end{cases}
\\ \text{as $N \to \infty$}.
\end{multline}
In other words: as the number of measurements becomes
large, the plausibility of the \so s that encode a
plausibility distribution not equal to one of the measured
frequencies vanishes, so that the whole plausibility gets
concentrated on the \so s encoding plausibility distributions
equal to the possible frequencies. This is an intuitively
satisfying result. The data single out a set of \so s, and
these are then given weight according to the prior
$\pf(\zrh \cond I)\, \di\zrh$, specified by us.

If $\zzfi$ degenerates into a single frequency value $\zfi$, the expression
above becomes, as shown in~\citep{portamana2007},
\begin{equation}
  \label{eq:post_N_delt}
  \pf[\zrh \cond  (\zf=\zfi) \land I]\, \di\zrh
\propto
\pf(\zrh \cond I)\, \delt[\zqq(\zrh)-\zfi]\,\di\zrh,
\end{equation}
which was also intuitively expected.

Note that if the prior density vanishes for such \so s as
are singled out by the data, then the equations above
become meaningless (no normalisation is possible),
revealing a contradiction between the prior knowledge and
the measurement data.

\subsection{Present case}
\label{sec:pres_case}

In the case of our study, the derivation above shows that,
as $N\to \infty$ and the triple of relative frequencies
$\zf \equiv (f_1, f_2, f_3) \defd (N_1, N_2,N_3)/N$ tends
to some value $\zfi$, the diagonal elements of the
assigned \so\ $\zrh_{D_N \land I}$ tend to
\begin{equation}
  \label{eq:lim_rho_inf}
  \pf(i \cond \zrh_{D_N \land I}) \equiv
  (\zrh_{D_N \land I})_{ii} \to \zffi_i \quad \text{as
    $N\to \infty$}.
\end{equation}
Combining this with the results of \sect~\ref{sec:symm}
concerning the off-diagonal elements, we find that the
assigned \so\ has in the limit the form
\begin{equation}
\label{eq:rhomatrix_inf}
\zrh_{D_\infty \land I}=
\begin{pmatrix}
                      \zffi_1
                    & 0
                    &0 \\
                     0  
                    & \zffi_2
                    & 0 \\
                     0
                    & 0
                   &  \zffi_3
               \end{pmatrix},
\end{equation} 
for both studied priors. This is again an expected
result. Only the diagonal elements of the \so\ are
affected by the data, and as the data amount increases it
overwhelms the prior information affecting the diagonal
elements. Both priors are moreover symmetric in respect of
the off-diagonal elements, that get thus a vanishing
average.

\section{Discussion and conclusions}
\label{sec:discussion}

Bayesian quantum-state assignment techniques have been
studied for some time now but, as far as we know, never
been applied to the whole set of \so s of systems with
more than two levels. And they have never been used for
state assignment in real cases. In this study we have
applied such methods to a three-level system, showing that
the numerical implementation is possible and simple in
principle. This paper should therefore not only be of
theoretical interest but also be of use to
experimentalists involved in state estimation. The time
required to obtain the numerical results was relatively
short in this three-level case, which involved an
eight-dimensional integration. Application to higher-level
systems should also be feasible, if one considers that
integrals involving hundreds of dimensions are computed in
financial, particle-physics, and image-processing problems
(see \eg\ the (somewhat dated)
refs.~\citep{james1980,stewart1983,lavalleetal1997,sloanetal1998,novak2000}).

Bayesian methods always take into account prior knowledge.
We have given examples of state-assignment in the case of
``vague'' prior knowledge, as well as in the case of a
kind of somehow better knowledge assigning higher
plausibility to \so s in the vicinity of a given pure one.
A comparison of the resulting \so s for the same kind of
data is quickly obtained by looking at
figs.~\ref{fig:Figure1} and \ref{fig:Figure5} (or at the
respective \so s in table~\ref{tab:SO}). It is clear that
when the available amount of data is small (as is the case
in those figures, which concern data with no or only one
measurement outcome), prior knowledge is very relevant.
Any practised experimentalist usually has some kinds of
prior knowledge in many experimental situations, which
arise from past experience with similar situations. With
some practice in ``translating'' these kinds of prior
knowledge into distribution functions, one could employ
small amounts of data in the most efficient way.

The generalisation of the present study to data involving
different kinds of measurement is straightforward. Of
course, in the general case one has to numerically
determine a greater number of parameters (the $L_j$) and
therefore compute a greater number of integrals. It would
also be interesting to look at the results for other kinds
of priors, in particular ``special'' priors like the Bures
one~\citep{slater1999b,byrdetal2001,slater2001,slater2001b,sommersetal2003}.
We found a particular non-trivial numerical relation,
\eqn~\eqref{eq:int_conv}, between the results obtained for
the constant prior; it would be interesting to know
whether it holds exactly.

In the next paper~\citep{maanssonetal2007} we shall give examples of
numerical quantum-state assignment for data consisting in average values
instead of absolute frequencies; and besides the two priors considered here
we shall employ another prior studied by Slater~\citep{slater1995b}.

\begin{acknowledgements}
  AM thanks Professor Anders Karlsson for encouragement.
  PM thanks Louise for continuous and invaluable support,
  and the kind staff of the KTH Biblioteket, the Forum
  biblioteket in particular, for their irreplaceable work.

  \emph{Post scriptum:} We cordially thank Paul B. Slater for pointing out
  to us the method of cylindrical algebraic decomposition, by which some of
  the integrals of this paper can be solved analytically, and for other
  important remarks.
\end{acknowledgements}
  
\appendix*

\section{Determination of $\zcset$}
\label{sec:det_C}

Any hyperplane tangent to (supporting) a convex set must
touch the latter on at least an extreme
point~\citep{valentine1964,gruenbaum1967_r2003,rockafellar1970,mcmullenetal1971,broendsted1983,webster1994}.
To determine the hyper-sides of the minimal hyper-box
$\zcset$ containing $\zxset$ we need therefore consider
only the maximal points of the latter --- \ie, the pure
states.

A generic ray of a three-dimensional complex Hilbert space
can be written as
\begin{gather}
  \label{eq:gen_ray}
 \ket{\zpu} =   \zpa \,\ket{1} 
+ \e^{-\I\zfb} \zpb \,\ket{2} 
+  \e^{-\I\zfc} \zpc \,\ket{3},
\\ \intertext{with}
0\le \zfb,\zfc \le 2\pu, \qquad\zpa, \zpb,\zpc \ge 0,
\quad\zpa^2 + \zpb^2 +\zpc^2 =1;\label{eq:ranges_pure}
\end{gather}
note that any two of the parameters $\zpa$, $\zpb$, $\zpc$
can be chosen independently in the range $\clcl{0,1}$.
The corresponding pure \so\ is
  \begin{equation}
    \label{eq:so_pure}
    \ketbra{\zpu}{\zpu} =  
    \begin{pmatrix}
      \zpa^2 
& \e^{-\I\zfb} \zpa\zpb 
 & \e^{-\I\zfc} \zpa\zpc
      \\
\e^{\I\zfb} \zpa\zpb 
& \zpb^2
& \e^{-\I(\zfb-\zfc)}\zpb\zpc
      \\
 \e^{\I\zfc} \zpa\zpc
&\e^{\I(\zfb-\zfc)}\zpb\zpc
&\zpc^2
    \end{pmatrix}.
  \end{equation}
  All pure states have this form, with the parameters in
  the ranges~\eqref{eq:ranges_pure}. Equating this
  expression with the one in terms of the Bloch-vector
  components $(\zla_i)$, \eqn~\eqref{eq:rhomatrix}, we
  obtain after some algebraic manipulation a parametric
  expression for the Bloch vectors of the pure states:
\begin{equation}
\label{eq:bloch_pure}
\begin{aligned}
\zla_1 &= 2  \zpa\zpb \cos\zfb, 
&
\zla_2 &= 2 \zpa\zpb  \sin\zfb,
\\
\zla_3 &=  \zpa^2-\zpb^2, 
&
\zla_4 &= 2 \zpa\zpc  \cos\zfc,
\\
\zla_5 &= 2\zpa\zpc  \sin\zfc ,
&
\zla_6 &= \zpb\zpc \cos(\zfb-\zfc),
\\
\zla_7 &= \zpb\zpc \sin(\zfb-\zfc) , 
&
\zla_8 &=  \sqrt{3} (\zpb^2 -1/3).
\end{aligned}
\end{equation}
These parametric equations define the four-dimensional
subset of the extreme points of $\zxset$. It takes little
effort to see that, as $\zpa$, $\zpb$, $\zpc$, $\zfb$, and
$\zfc$ vary in the ranges~\eqref{eq:ranges_pure}, each of
the first seven coordinates above ranges in the interval
$\clcl{-1,1}$ and the eighth in the interval
$\clcl{-2/\sqrt{3}, 1/\sqrt{3}}$. The rectangular region
given by the Cartesian product of these intervals is thus
$\zcset$ as defined in \eqn~\eqref{eq:def_C}, \qed


\renewcommand{\bibpreamble}{Note: \texttt{arxiv} eprints are
  located at \url{http://arxiv.org/}.
}

\setlength{\bibsep}{0pt}


\bibliography{bibliography}

\begin{figure*}[ph!]
\includegraphics[width=\textwidth]
{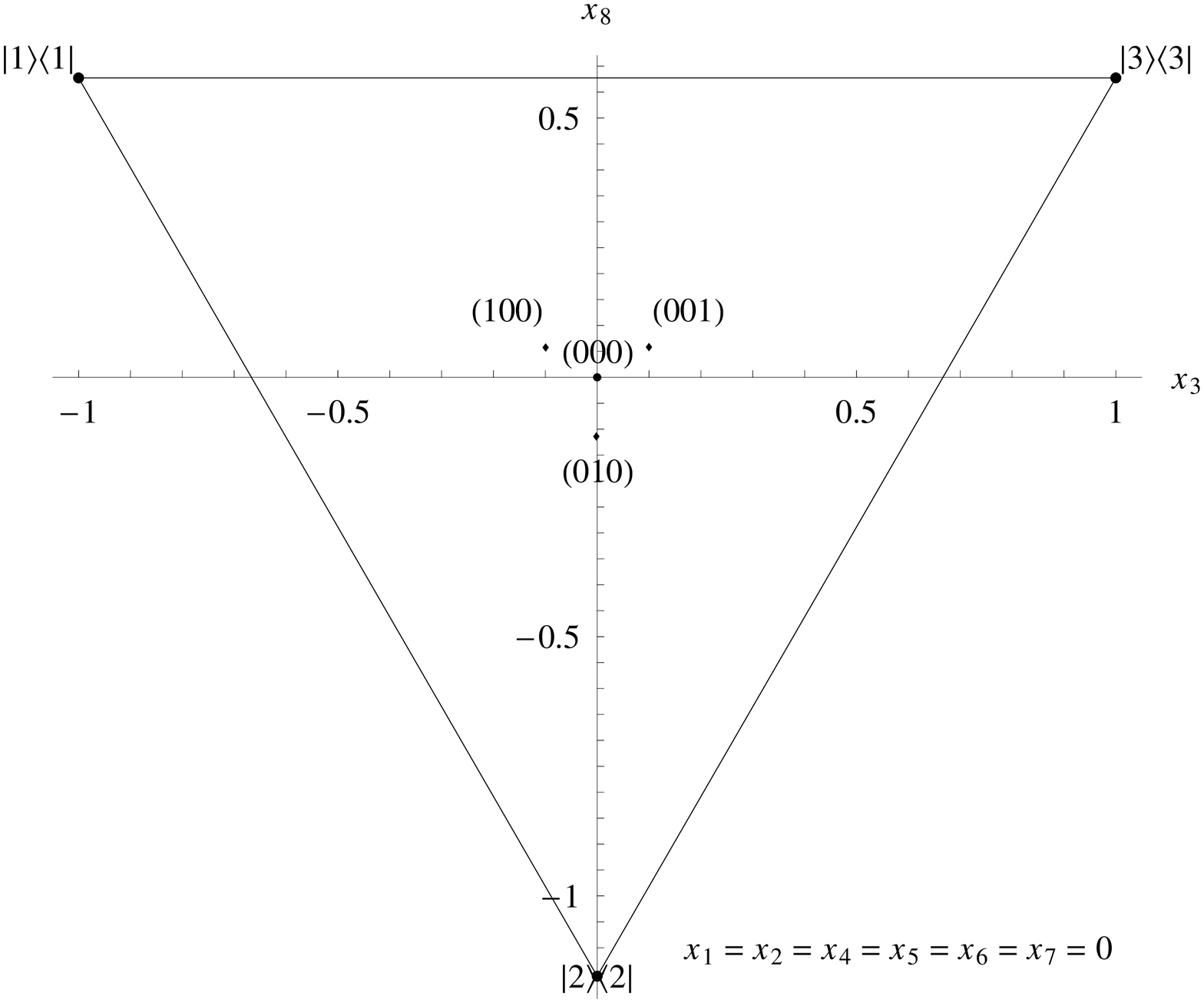}
\caption{Bloch vectors of the assigned \so\ for prior
  knowledge $\zI$ and
  absolute-frequency triples with $N=0$ and $N=1$,
  computed by numerical integration.
The large triangle in the figures is the two-dimensional
section of the set $\zxset$ along the plane
$\origo\zla_3\zla_8$. The numerical-integration
uncertainty in the $\zla_3$ and $\zla_8$ components is
$\pm 0.005$. In the case of no data ($N=0$), the \so\ assigned
on the basis of the prior knowledge $\zI$ alone is the
``completely mixed'' one $\bm{I}_3/3$.  Note that that all
  the components of all four
  points have been determined by numerical integration,
  even those that can be exactly determined by symmetry
  arguments. Within the given uncertainties, numerical computations yielded the exact
  results.
  \label{fig:Figure1}}
\end{figure*}

\begin{figure*}[ph!]
\includegraphics[width=\textwidth]
{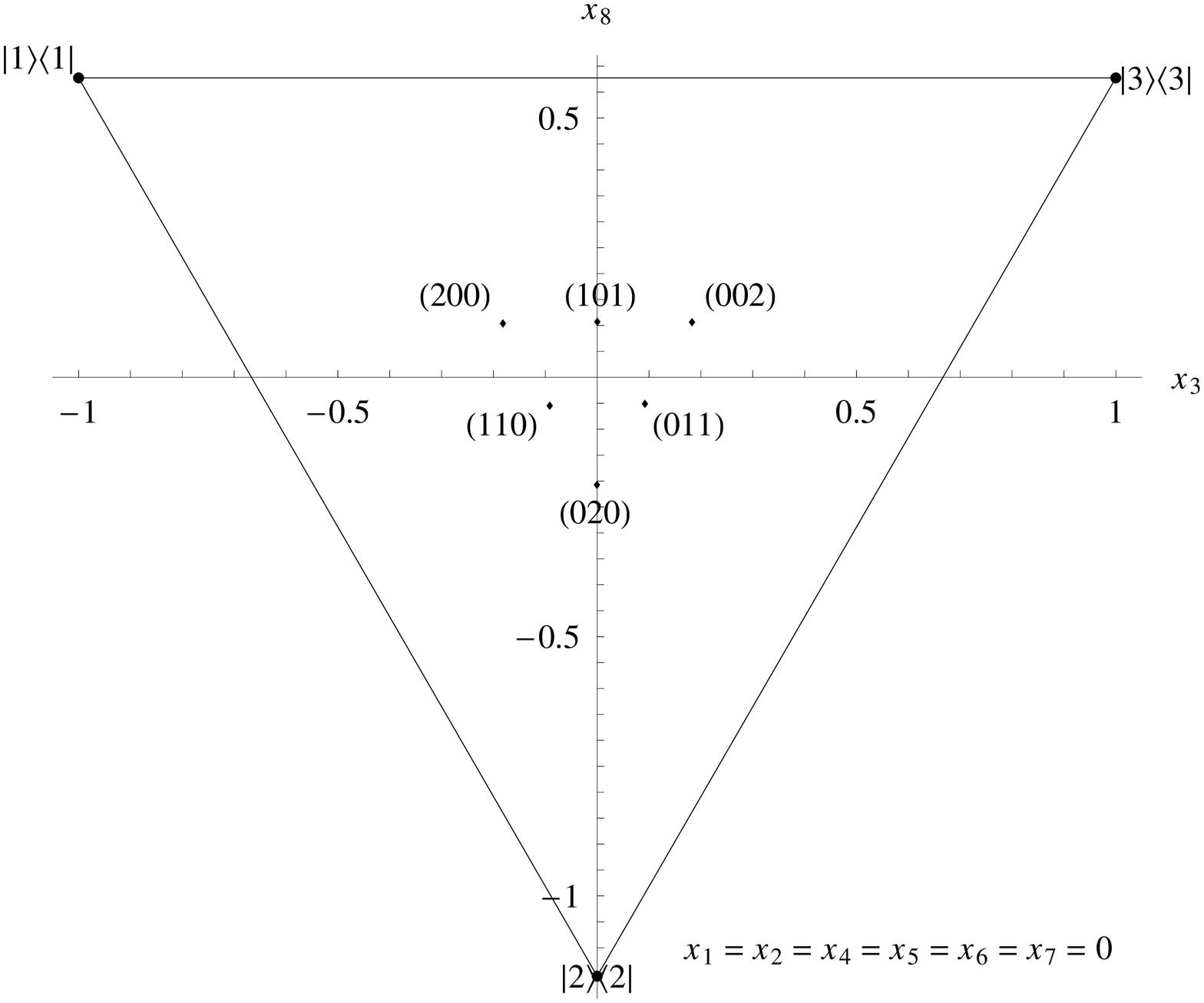}
\caption{Bloch vectors of the assigned \so\ for prior
  knowledge $\zI$ and
  absolute-frequency triples with $N=2$,
  computed by numerical integration.
The large triangle in the figures is the two-dimensional
section of the set $\zxset$ along the plane
$\origo\zla_3\zla_8$. The numerical-integration
uncertainty in the $\zla_3$ and $\zla_8$ components is
$\pm 0.0025$.  Note that that all
  the components of all six
  points have been determined by numerical integration,
  even those that can be exactly determined by symmetry
  arguments. Within the given uncertainties, numerical computations yielded the exact
  results.
  \label{fig:Figure2}}
\end{figure*}

\begin{figure*}[ph!]
\includegraphics[width=\textwidth]
{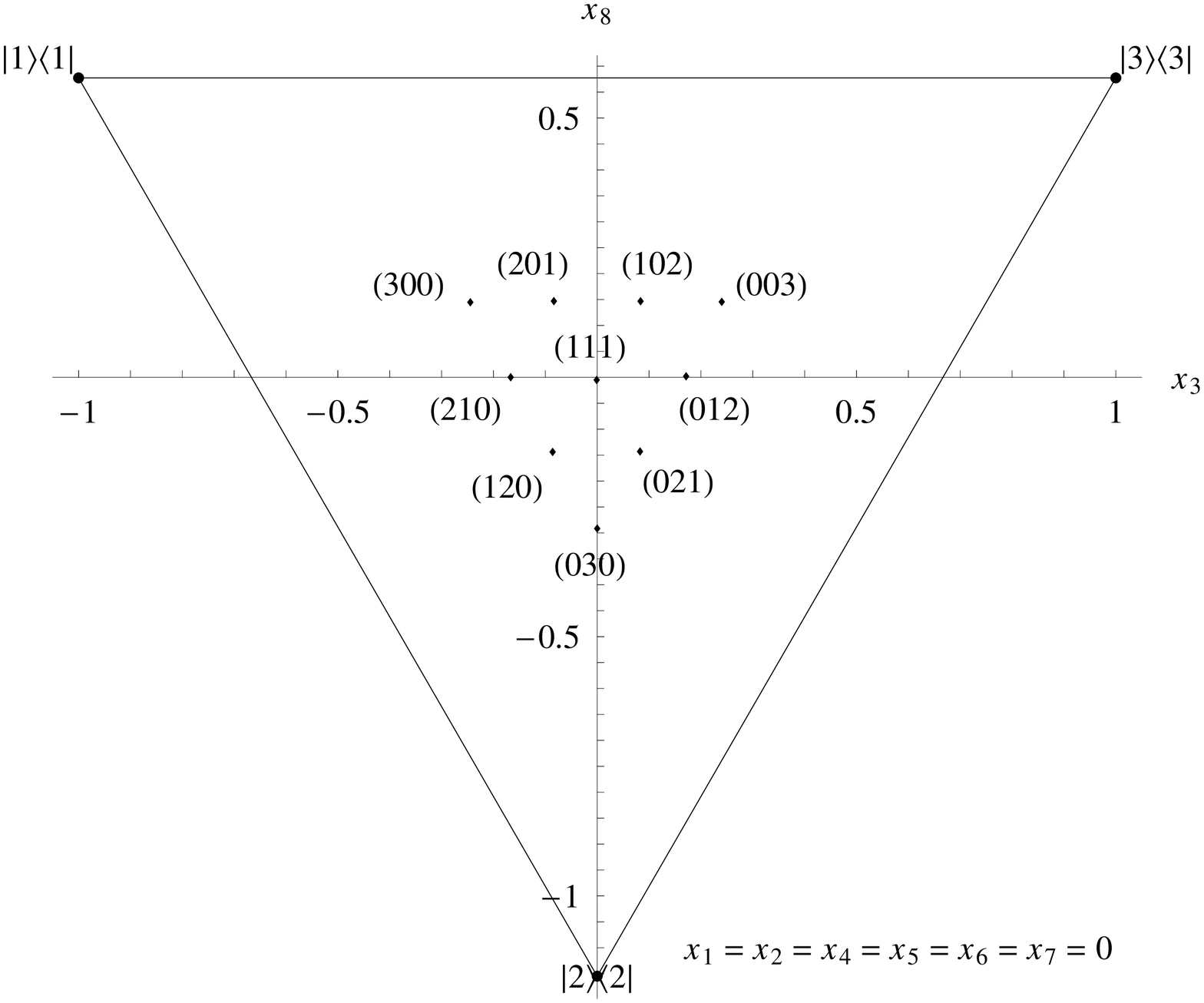}
\caption{Bloch vectors of the assigned \so\ for prior
  knowledge $\zI$ and
  absolute-frequency triples with $N=3$,
  computed by numerical integration.
The large triangle in the figures is the two-dimensional
section of the set $\zxset$ along the plane
$\origo\zla_3\zla_8$. The numerical-integration
uncertainty in the $\zla_3$ and $\zla_8$ components is
$\pm 0.005$.  Note that that all
  the components of all ten
  points have been determined by numerical integration,
  even those that can be exactly determined by symmetry
  arguments. Within the given uncertainties, numerical computations yielded the exact
  results.
  \label{fig:Figure3}}
\end{figure*}

\begin{figure*}[ph!]
\includegraphics[width=\textwidth]
{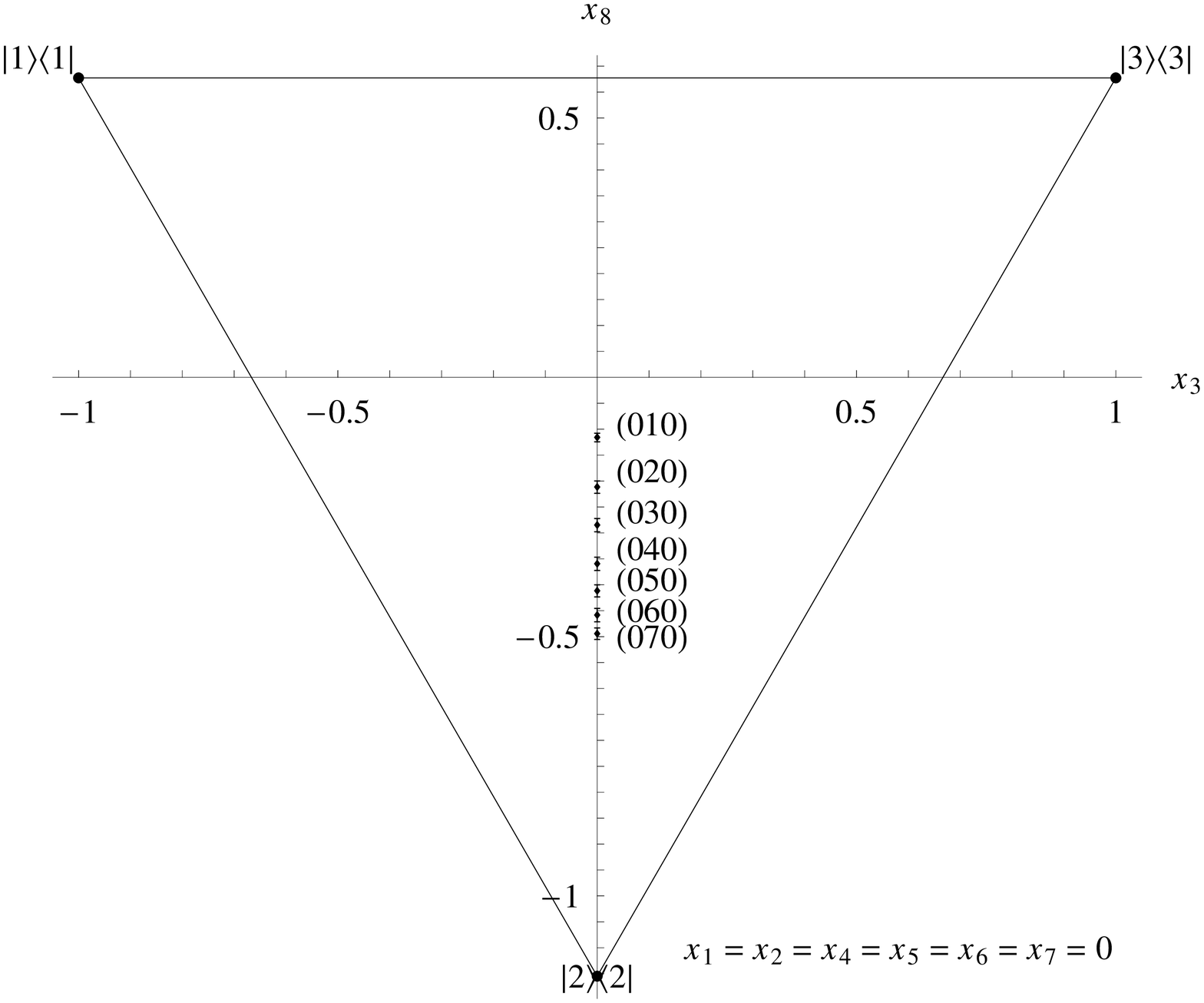}
\caption{Bloch vectors of the assigned \so\ for prior
  knowledge $\zI$ and
  absolute-frequency triples of the form $(0,N,0)$, with
  $N=1,2,3,4,5,6,7$,
  computed by numerical integration.
The large triangle in the figures is the two-dimensional
section of the set $\zxset$ along the plane
$\origo\zla_3\zla_8$. The numerical-integration
uncertainty in the $\zla_3$ and $\zla_8$ components is
$\pm 0.015$.  Only the $\zla_8$ component was determined
by numerical integration; the $\zla_3$ vanishes for symmetry reasons.  \label{fig:Figure4}}
\end{figure*}

\begin{figure*}[ph!]
\includegraphics[width=\textwidth]
{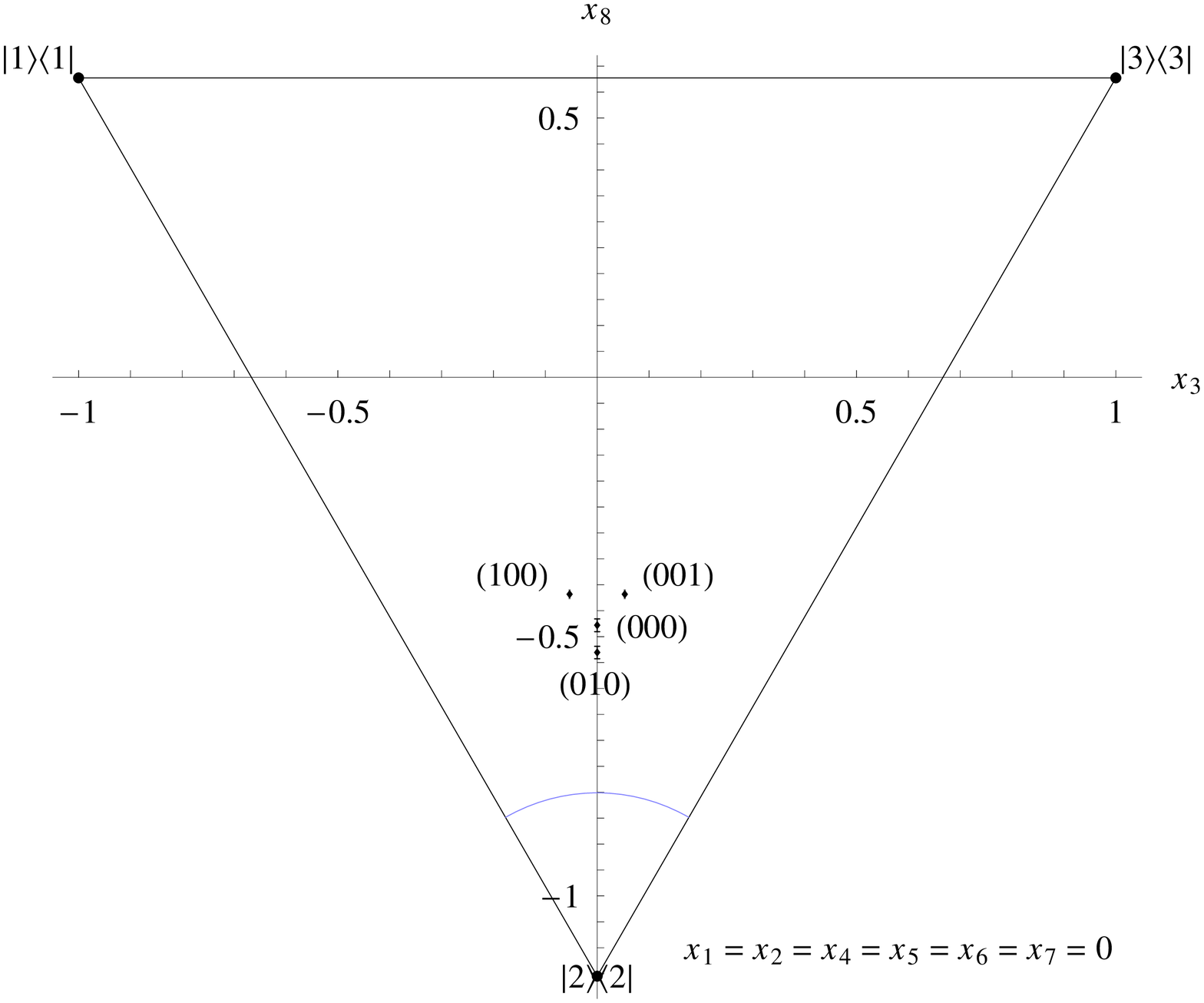}
\caption{Bloch vectors of the assigned \so\ for prior
  knowledge $\zJ$ and absolute-frequency triples with
  $N=0$ and $N=1$, computed by numerical integration. The
  large triangle in the figures is the two-dimensional
  section of the set $\zxset$ along the plane
  $\origo\zla_3\zla_8$. The prior knowledge is represented
  by a Gaussian-like distribution of ``breadth''
  $s=1/(2\sqrt(2)$ centred on the pure \so\ $\ztwo$; see
  \sect~\ref{sec:prior}. The small circular arc is the
  locus of the Bloch vectors (on the plane) at a distance
  $\abs{\zll - \zllc} =s$ from the vector $\zllc \defd
  (0,0,0,0,0,0,0,-2/\sqrt{3})$ corresponding to the \so\
  $\ztwo$. In the case of no data ($N=0$), the \so\ assigned
on the basis of the prior knowledge $\zJ$ alone lies in between
the completely mixed one and the pure one $\ztwo$.
\label{fig:Figure5}}
\end{figure*}


\end{document}